\documentclass[fleqn,usenatbib]{mnras}

\usepackage{newtxtext,newtxmath}

\usepackage[T1]{fontenc}

\DeclareRobustCommand{\VAN}[3]{#2}
\let\VANthebibliography\thebibliography
\def\thebibliography{\DeclareRobustCommand{\VAN}[3]{##3}\VANthebibliography}

\usepackage{graphicx}	
\usepackage{amsmath}	

\usepackage{subcaption}

\title[Towards inferring the geometry of kilonovae]{Towards inferring the geometry of kilonovae}

\author[C. E. Collins et al.]{Christine E. Collins,$^{1}$\thanks{E-mail: c.collins@gsi.de}
Luke J. Shingles,$^{1}$
Andreas Bauswein,$^{1}$
Stuart A. Sim,$^{2}$
Theodoros Soultanis,$^{1}$\newauthor
Vimal Vijayan,$^{1,3}$
Andreas Flörs,$^{1}$
Oliver Just,$^{1,4}$
Gerrit Leck,$^{1,5}$
Georgios Lioutas,$^{1}$
Gabriel Mart\'inez-Pinedo,$^{1,5}$\newauthor
Albert Sneppen,$^{6,7}$
Darach Watson$^{6,7}$ and
Zewei Xiong$^{1}$
\\
    $^{1}$GSI Helmholtzzentrum f\"{u}r Schwerionenforschung, Planckstraße 1, 64291 Darmstadt, Germany\\
    $^{2}$Astrophysics Research Centre, School of Mathematics and Physics, Queens University Belfast, Belfast BT7 1NN, UK\\
    $^{3}$Department of Physics and Astronomy, Ruprecht-Karls-Universität Heidelberg, Im Neuenheimer feld 226, 69120 Heidelberg, Germany\\
    $^{4}$Astrophysical Big Bang Laboratory, RIKEN Cluster for Pioneering Research, 2-1 Hirosawa, Wako, Saitama 351-0198, Japan\\
    $^{5}$Institut {f\"ur} Kernphysik (Theoriezentrum), Technische Universit{\"a}t Darmstadt, Schlossgartenstra{\ss}e 2, D-64289 Darmstadt, Germany\\
    $^{6}$Cosmic Dawn Center (DAWN)\\
    $^{7}$Niels Bohr Institute, University of Copenhagen, Jagtvej 128, København 2200, Denmark\\
}

\date{Accepted XXX. Received YYY; in original form ZZZ}

\pubyear{2024}

\begin{document}
\label{firstpage}
\pagerange{\pageref{firstpage}--\pageref{lastpage}}
\maketitle

\begin{abstract}
    Recent analysis of the kilonova, AT2017gfo, has indicated that this event was highly spherical.
    This may challenge hydrodynamics simulations of binary neutron star mergers, which usually predict a range of asymmetries, and radiative transfer simulations show a strong direction dependence.
    Here we investigate whether the synthetic spectra from a 3D kilonova simulation of asymmetric ejecta from a
    hydrodynamical merger simulation can be compatible with the observational constraints suggesting a high degree of sphericity in AT2017gfo.
    Specifically, we determine whether fitting a simple P-Cygni line profile model leads to a value for the photospheric
    velocity that is consistent with the value obtained from the expanding photosphere method.
    We would infer that our kilonova simulation is highly spherical
    at early times, when the spectra resemble a blackbody distribution.
    The two independently inferred photospheric velocities can be very similar, implying a high degree of sphericity,
    which can
    be as spherical as inferred for AT2017gfo, demonstrating that the photosphere can appear spherical even for asymmetrical ejecta.
    The last-interaction velocities of radiation escaping the simulation show a high degree of sphericity, supporting the inferred symmetry of the photosphere.
    We find that when the synthetic spectra resemble a blackbody the expanding photosphere method can be used to obtain an accurate luminosity distance (within 4 -- 7 per cent).
\end{abstract}

\begin{keywords}
neutron star mergers -- radiative transfer -- methods: numerical
\end{keywords}



\section{Introduction}

The kilonova AT2017gfo, which was coincident with the gravitational wave signal GW170817 \citep{abbott2017a},
has provided us with a wealth of observations \citep[e.g.,][]{smartt2017a, pian2017a, villar2017a, coulter2017a}.
Recently, \citet{sneppen2023a} presented evidence that AT2017gfo was highly spherical, which is surprising given that binary neutron star merger simulations show strong asymmetries \citep{bauswein2013a, sekiguchi2015a, bovard2017a, radice2018a, foucart2023a, combi2023a}.

\citet{sneppen2023a} inferred the expansion velocity of the ejecta by analysing the most prominent feature in the
spectra of AT2017gfo, which has been suggested to be a \ion{Sr}{II} P-Cygni feature
(\citealt{watson2019a, gillanders2022a, domoto2021a, domoto2022a}, although see \citealt{tarumi2023a}
for discussion of \ion{He}{I} as an alternative explanation for this feature). Such line profile analysis is predominantly sensitive to the line of sight velocity component.
\citet{sneppen2023a} also inferred a photospheric radius using the 
expanding photosphere method (EPM; \citealt{baade1926a, kirshner1974a, eastman1996a})
and, assuming homologous expansion of the ejecta, converted this to an expansion velocity of the photosphere. 
This method is primarily sensitive to the expansion velocity perpendicular to the line of sight.
They found remarkable consistency between the velocities obtained via these two methods across two observed epochs, suggesting consistency between line of sight and perpendicular expansion velocities, which is indicative of a near-spherical explosion.
To quantify the inferred sphericity, \citet{sneppen2023a} defined a zero-centred asymmetry index, $\Upsilon=\frac{v_{\perp}-v_{\|}}{v_{\perp}+v_{\|}}$, where $v_{\|}$ is the expansion velocity of the photosphere primarily along the line of sight (obtained from the P-Cygni profile analysis) and $v_{\perp}$ is the expansion velocity of the photosphere primarily in the direction perpendicular to the line of sight (obtained from the EPM), and found $\Upsilon=0.00 \pm 0.02$\footnote{
This value of $\Upsilon$ was found using the Hubble expansion parameter, H$_0$, as suggested from the cosmic microwave background (CMB; \citealt{planckcollaboration2020a}) to infer v$_\perp$. Using the value inferred from the local distance ladder \citep{riess2022a} an asymmetry of $\Upsilon = -0.04 \pm 0.03$ was derived.
}.
In addition to this analysis, they identify that the shape of the P-Cygni feature is best matched
by assuming a spherical photosphere across multiple epochs (see \citealt{sneppen2023a}).

\citet{shingles2023a} presented a three-dimensional radiative transfer calculation using ejecta from a binary neutron star merger simulation coupled to a nucleosynthesis network. 
In the polar directions, the spectra resembled the observations of AT2017gfo remarkably well, considering the model was not tuned in any way to match this event, although at earlier times than those observed (see \citealt{shingles2023a} for discussion).
The synthetic observables showed variations with both polar and azimuthal viewing angles,
and therefore are not isotropic.
Here, we analyse this simulation to determine whether we can infer information about the underlying symmetry of the ejecta from the synthetic observables.

The aim of this paper is to determine whether the geometry and overall degree of sphericity of the ejecta is encoded in the observables as suggested by \citet{sneppen2023a}. To this end, we consider the synthetic observables predicted by the 3D radiative transfer calculation by \citet{shingles2023a} from an asymmetric merger simulation to analyse the velocities that would be derived from these observables following the method laid out by \citet{sneppen2023a},
and investigate whether the apparent consistency of $v_\|$ and $v_\perp$ is a fundamental challenge to the explosion model.
Determining whether the geometry of simulations is compatible with observations
will further our understanding of binary neutron star mergers, including the high density Equation of State, the dynamics of matter ejection, including the role of neutrinos, and the underlying rapid neutron capture (r-process) nucleosynthesis.

The EPM is typically used to measure luminosity distances, D$_\mathrm{L}$, to supernovae, and 
can be used to obtain distance measurements
independently of the cosmological distance ladder,
which can lead to independent measurements of the Hubble constant \citep{dejaeger2017a, gall2018a, sneppen2023b, sneppen2023a}.
\citet{sneppen2023a} applied the EPM to measure the distance to AT2017gfo and found D$_\mathrm{L}$ to be consistent with previous distance measurements.
In addition to inferring the geometry of our simulations, we aim to test how accurately the luminosity distance to our synthetic spectra can be measured using the EPM.

\section{Simulations}

\subsection{Merger ejecta symmetry}

\begin{figure}
\centering
\includegraphics[width=0.4\textwidth]{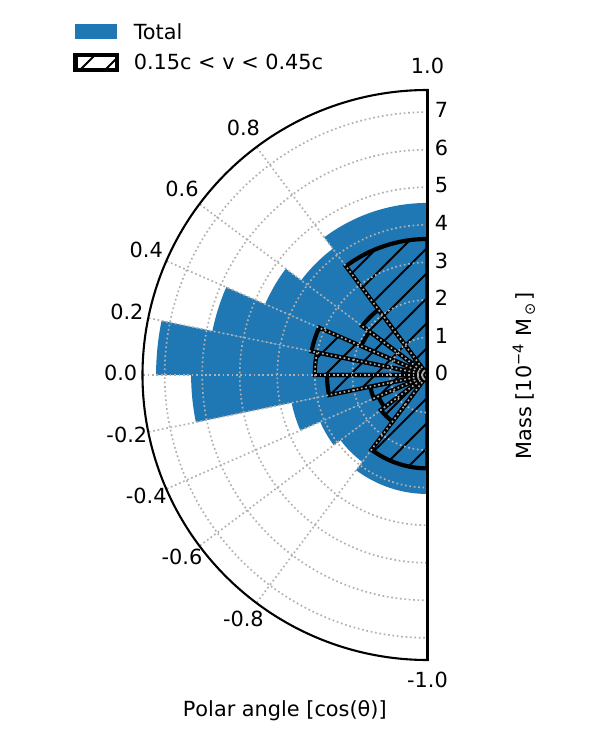}

\caption{Mass ejected by polar angle, where each bin has an equal solid-angle.
The height of each bar indicates the mass ejected within the solid-angle bin.
The black lines indicate the mass ejected into the velocity range which is approximately in the line forming region of the kilonova within the first day (assuming homologous expansion).
}

\label{fig:massdistribution}
\end{figure}

 \begin{table*}
     \centering
         \caption{Mass ejected within equal solid-angle bins, defined by polar angle, at the poles and around the equator.
         We define $\mu_\mathrm{ej}$ to be the range in polar-angle of each bin in the ejecta, which each have a width of cos$(\theta)=0.2$.
         Included is the total mass, M, within the solid-angle and the mass, M$_v$, within the velocity range
             (assuming homologous expansion) 0.15c < $v$ < 0.45c, which is approximately
             within the spectral line-forming region during the first day of the kilonova.
             This is also shown for the mass of Sr, M$^\mathrm{Sr}$, and the mass of Sr within the velocity range, M$_{v}^\mathrm{Sr}$.}
 \begin{tabular}{ccccc}
     \hline
    $\mu_\mathrm{ej}$                 & M                     & M$_v$ (0.15c < $v$ < 0.45c)& M$^\mathrm{Sr}$     & M$_{v}^\mathrm{Sr}$ (0.15c < $v$ < 0.45c) \\
    {[cos($\mathrm {\theta }$)]} & [10$^{-4}$ M$_\odot$] & [10$^{-4}$ M$_\odot$]      & [10$^{-4}$ M$_\odot$] & [10$^{-4}$ M$_\odot$] \\ \hline
     {[$0.8, 1.0$]} ($+$ve pole)    & 4.58                  & 3.62                       & 0.46                  & 0.38 \\
    {[$-1.0, -0.8$]} ($-$ve pole)  & 3.17                  & 2.49                       & 0.45                  & 0.36 \\
     {[$0.0, 0.2$]} (eq.)          & 7.23                  & 3.00                       & 0.24                  & 0.065 \\
    {[$-0.2, 0.0$]} (eq.)         & 6.30                  & 2.68                       & 0.25                  & 0.074 \\
     \hline
     \end{tabular}
     \label{tab:mass}
 \end{table*}

 \begin{table*}
     \centering
         \caption{Asymmetry parameter comparing mass ejected within equal solid-angle bins near the equator and at the poles using the values in Table~\ref{tab:mass}.
         The range in polar-angle of each bin, $\mu_\mathrm{ej}$, (with a width of cos$(\theta)=0.2$) is listed.
         }
 \begin{tabular}{cccccc}
     \hline
     $\mu_\mathrm{ej}$ (pole)          & $\mu_\mathrm{ej}$ (equator)     & $\Upsilon_\mathrm{M}$ & $\Upsilon_{\mathrm{M},v}$ & $\Upsilon_{\mathrm{M}}^\mathrm{Sr}$  & $\Upsilon_{\mathrm{M},v}^\mathrm{Sr}$ \\
     {[cos($\mathrm {\theta }$)]} & [cos($\mathrm {\theta }$)] &                       &                &                             &                 \\ \hline
     {[$0.8, 1.0$]} ($+$ve pole)   & [$0.0, 0.2$] (eq.)           &  $0.22$                 & $-0.094$         &  $-0.31$                     & $-0.71$            \\
     {[$-1.0, -0.8$]} ($-$ve pole) & [$-0.2, 0.0$] (eq.)          &  $0.33$                 & $0.037$          &  $-0.29$                      & $-0.66$           \\
     \hline
     \end{tabular}
     \label{tab:asymm-mass}
 \end{table*}

\begin{figure*}
\centering
\subcaptionbox{Total mass}{\includegraphics[width=0.4\textwidth]{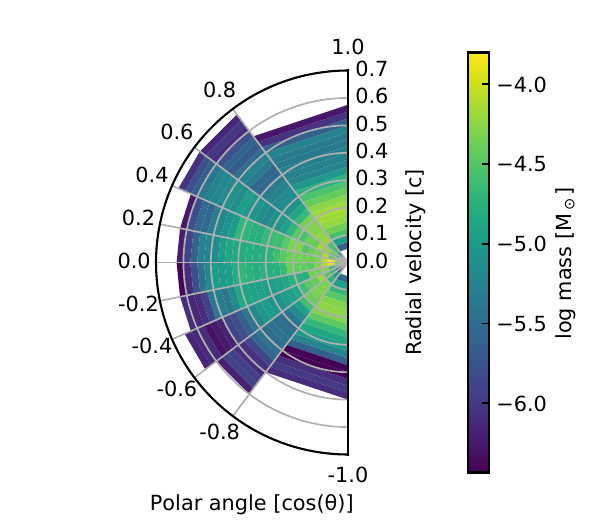}}
\subcaptionbox{Sr mass}{\includegraphics[width=0.4\textwidth]{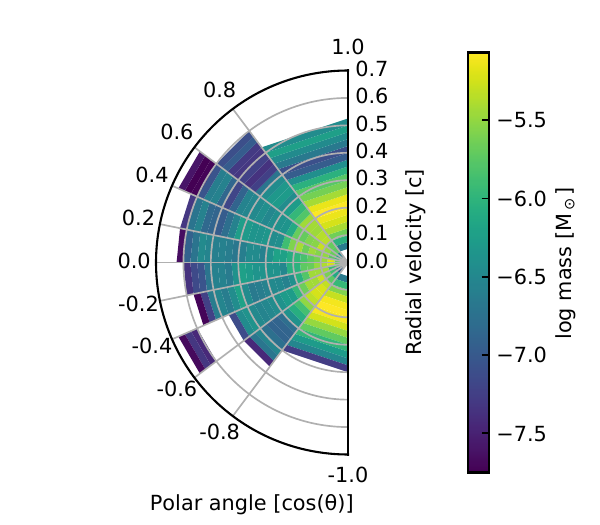}}

\caption{
Total mass (a) and mass of Sr (b)
ejected into polar angle bins, where each angle bin has an equal solid-angle with a width of cos($\theta$) = 0.2.
The colour scale indicates the mass lying within each radial zone.
Note a lower cut has been placed on mass to highlight the model structure.
}

\label{fig:massdistribution-Sr}
\end{figure*}

We consider the 3D binary neutron star merger simulation of equal mass 1.35 M$_\odot$ neutron stars used by \citet{shingles2023a} and \citet{collins2023a}.
{This merger simulation was carried
out with a 3D general relativistic smooth particle hydrodynamics
(SPH) code \citep{oechslin2002a, bauswein2013a}, which adopts the conformal flatness condition on the spatial metric \citep{isenberg1980a, wilson1996a} and includes an advanced neutrino leakage treatment \citep{ardevol2019a}. We employ the SFHo equation of state \citep{steiner2013a}.
As discussed by \citet{collins2023a}, the ejecta from this merger simulation show asymmetries both in the density structure and in the distribution of $Y_e$ (see figures 1 and 2 in \citealt{collins2023a}).
A higher $Y_e$ is found near the poles, primarily due to the inclusion of a neutrino treatment, and a lower $Y_e$ is found near the equator.
Nuclear network calculations were carried out for each SPH particle trajectory (see \citealt{collins2023a} for details of the calculation) to model the spatial distribution and time dependence of the early heating and elemental composition.
The evolution of the composition and energy released due to radioactive decays is then followed by \textsc{artis} (see \citealt{shingles2023a} for details).
}
The mass ejected into equal solid-angle bins, spaced by polar angle, for this model is shown in Figure~\ref{fig:massdistribution}.
The simulation predicts more mass per solid angle near the equator compared to near the poles.
There is a mild equatorial (i.e., north vs. south) asymmetry, visible in
Figure~\ref{fig:massdistribution}, which to some degree may be physical (as a result of stochastic fluctuations and hydrodynamic instabilities in the matter flow), but may also be amplified by numerical effects (e.g. particle noise).
We discuss these aspects in more detail in Appendix~\ref{sec:deviations_from_symm}.
We note that the merger simulation
{provides only dynamical ejecta (evolved until 20 ms after the merger).}
The merger simulation shows some variation in asymmetry with time, and it terminates before the matter ejection ceases and the ejecta configuration reaches its final state. 
We thus do not expect an overall match with AT2017gfo.
{The impact of the continued mass ejection (secular ejecta) on our simulation results should be investigated in future.
However, given that this dynamical ejecta model is able to produce spectra comparable to those observed for AT2017gfo (see \citealt{shingles2023a} and discussion in Section~\ref{sec:synthetic_obs}) we use this ejecta model as the basis for our analysis.}

Also shown is the mass of material ejected in the velocity range 0.15c < $v$ < 0.45c, M$_v$, which is approximately in the line forming region during the first day of the kilonova (Section~\ref{sec:comparelineveltosimulated}).
Note that 44~per~cent of the ejecta mass is below 0.15c.
Figure~\ref{fig:massdistribution-Sr} shows how the ejecta mass is distributed with polar angle and radial velocity {(see figure 1 in \citealt{collins2023a} for a 3D rendering of the ejecta structure).}
The ejecta exhibit asymmetry both in the total mass ejected into different directions, and in the velocity at which the bulk of the mass is ejected.

To quantify the asymmetry of the ejecta in the simulation,
we define a zero-centred asymmetry parameter\footnote{
A zero-centred asymmetry parameter was defined by \citet{sneppen2023a} to quantify the sphericity in the inferred velocities of the photosphere.
Similarly, we use zero-centred asymmetry parameters to quantify sphericity, however the measurements on the sphericity of the mass and luminosity are not directly comparable to the quantity inferred by \citet{sneppen2023a}.}
for the ejected mass,
\begin{equation}
    \Upsilon_{\mathrm{M}}=\mathrm{\frac{{M_{eq}}-M_{pole}}{M_{eq}+M_{pole}}},
\end{equation}
where $\mathrm{M_{eq}}$ is the mass within a solid-angle at the equator and $\mathrm{M_{pole}}$ is the mass within an equal solid-angle at the pole.
The masses within equal solid-angles (plotted in Figures~\ref{fig:massdistribution} and \ref{fig:massdistribution-Sr}) are listed in Table~\ref{tab:mass} for solid-angles near the poles and equator.
The values of $\Upsilon_{\mathrm{M}}$ are listed in Table~\ref{tab:asymm-mass}.

The values of $\Upsilon_{\mathrm{M}}$ indicate a moderate level of asymmetry in the mass per solid-angle ejected near the poles compared to the equator, however, $\Upsilon_{\mathrm{M},v}$ (considering only the mass ejected within the velocity range 0.15c < $v$ < 0.45c) indicates that the mass approximately within the line forming region within the first day of the kilonova is more symmetric,
which may lead to a higher level of symmetry in the synthetic observables than indicated by $\Upsilon_{\mathrm{M}}$.

The abundances of synthesised elements in the ejecta also show asymmetries.
We show the distribution of Sr
synthesised in the ejecta in Figure~\ref{fig:massdistribution-Sr}, which is predominantly responsible for the strongest feature in our simulated spectra \citep{shingles2023a}.
Note that the feature is also composed of contributions from Y and Zr, which show a similar distribution in the ejecta to Sr.
We show the total mass of these and other representative elements
ejected by polar angle in Figure~\ref{fig:mass-histograms-elements} {in Appendix~\ref{sec:supplementalfigs}}.
Higher masses of Sr, Y and Zr are synthesised near the poles than near the equator (due to the higher $Y_e$ near the poles, {see figure 2 of \citealt{collins2023a}}).
\citet{sneppen2023a} suggest that the line shape of the spectral feature in AT2017gfo indicates a near spherical distribution of Sr, which is not shown by the distribution of Sr synthesised in our model.
Using the asymmetry parameter, $\Upsilon_{\mathrm{M}}^\mathrm{Sr}$, listed in Table~\ref{tab:asymm-mass}, the mass distribution of Sr shows a moderate level of asymmetry.
Within the velocity range 0.15c < $v$ < 0.45c, the distribution of Sr with polar angle is very asymmetric (e.g.,~$\Upsilon_{\mathrm{M,}v}^\mathrm{Sr}=-0.71$).

\subsection{Radiative transfer simulation}

Given the asymmetry of the ejecta model, we aim to test whether this level of asymmetry can be consistent with the observational constraints inferred by \citet{sneppen2023a} for AT2017gfo,
which appear to favour near spherical symmetry.
For this,
we analyse the 3D radiative transfer kilonova simulation, 3D~AD2, carried out by \citet{shingles2023a} to identify whether,
if subjected to a similar analysis as data from actual observations, it can match (or be ruled out by) constraints on its apparent sphericity, such as the $\Upsilon$ parameter used by \citet{sneppen2023a}.
This simulation was carried out using the multi-dimensional, time-dependent Monte Carlo radiative transfer code \textsc{artis} (\citealt{sim2007b, kromer2010a, shingles2020a, shingles2023a}, based on the methods of \citealt{lucy2002a, lucy2003a, lucy2005a}).
We note that in \textsc{artis} there is no photosphere defined in the simulation.
We can therefore infer the apparent photosphere from the synthetic observables (using the same methods as for observations) without the constraint of a photospheric boundary being imposed in the simulation.
To reduce Monte Carlo noise in the orientation-dependent synthetic observables, escaping packets of radiation are binned into ten equal solid-angle bins, defined by polar angle.

\section{Results}

\subsection{Synthetic observables}
\label{sec:synthetic_obs}
The synthetic light curves and spectra for this simulation are presented by \citet{shingles2023a}.
{The simulated light curves were fainter than observed for AT2017gfo, likely due to the lower ejecta mass in the model (0.005 M$_\odot$) than that inferred for AT2017gfo ($0.04 \pm 0.01$ M$_\odot$ was estimated by \citealt{smartt2017a}).
The spectral evolution in the polar direction was similar to that observed in AT2017gfo, showing a prominent feature primarily shaped by \ion{Sr}{II}, however, the simulated spectra show a more rapid evolution than observed in AT2017gfo.
}
In this section, we focus on the observer orientation dependence and the level of isotropy shown by the synthetic observables.

\subsubsection{Observer orientation variation in bolometric luminosity}

\begin{figure}

\includegraphics[width=0.48\textwidth]{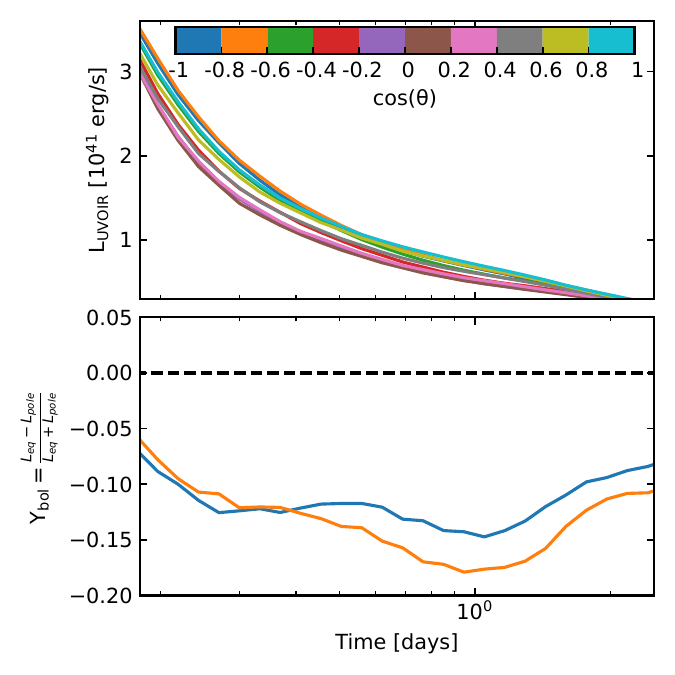}

\caption{Bolometric light curves averaged over azimuthal angle, showing the variation with polar angle.
Also shown is the zero-centred asymmetry index,
$\Upsilon_{\mathrm{bol}}$
comparing the luminosity at the poles to the luminosity at the equator.
In the lower plot, the blue line shows the negative pole compared to the equator ($-1 <$ cos($\mathrm{\theta}$) $< -0.8$ and \mbox{$-0.2 <$ cos($\mathrm{\theta}$) $< 0$})
and orange shows the positive pole compared to the equator  ($0.8 <$ cos($\mathrm{\theta}$) $< 1$ and \mbox{$0 <$ cos($\mathrm{\theta}$) $< 0.2$}).
}

\label{fig:bollightcurves}
\end{figure}

As a first quantification of the degree to which the simulation predicts observer orientation dependencies we compare the bolometric light curves in different directions.
As presented by \citet{shingles2023a}, the light curves in the lines of sight at the poles are brighter than those at the equator, due to lower densities and lower $Y_e$ near the poles (see \citealt{collins2023a}).
In Figure~\ref{fig:bollightcurves} we plot the angle-dependent bolometric light curves as
isotropic-equivalent luminosities (i.e.,
from the simulation we record the energy emitted per second into each
solid angle bin to obtain light curves in erg~s$^{-1}$~sr$^{-1}$ for each orientation;
we then scale these to an equivalent isotropic luminosity by multiplication
by the full-sphere solid angle of $4\pi \cdot$sr).
Note that the solid-angle bins around the poles encompass the inferred observer angle of AT2017gfo (between 19$^{\circ}$ and 25$^{\circ}$; \citealt{mooley2022a}). 

We define a zero-centred asymmetry index
\begin{equation}
    \Upsilon_{\mathrm{bol}}=\mathrm{\frac{L_{eq}-L_{pole}}{L_{eq}+L_{pole}}}
\end{equation}
 (similar to \citealt{sneppen2023a}, but not directly comparable), 
where $\mathrm{L_{pole}}$ is the luminosity at either pole and $\mathrm{L_{eq}}$ is the luminosity near the equator.
The luminosity emitted into the solid-angle bins above ($0 <$ cos($\mathrm{\theta}$) $< 0.2$) and below ($-0.2 <$ cos($\mathrm{\theta}$) $< 0$) the equator is almost identical -- see Figure~\ref{fig:bollightcurves}.
The maximum deviation of $\Upsilon_{\mathrm{bol}}$ between the poles and the equator is $\Upsilon_{\mathrm{bol}} \approx -0.18$ (Figure~\ref{fig:bollightcurves}).
This clearly demonstrates that the synthetic observables for this ejecta model are not isotropic,
however,
according to this simple asymmetry metric, the synthetic observables are not as asymmetric as the mass distribution of the ejecta.
We discuss this further in Section~\ref{sec:band_lightcurves}.

\subsubsection{Observer orientation variation in band-limited light curves}
\label{sec:band_lightcurves}

\begin{figure*}
\centering
\includegraphics[width=0.8\textwidth]{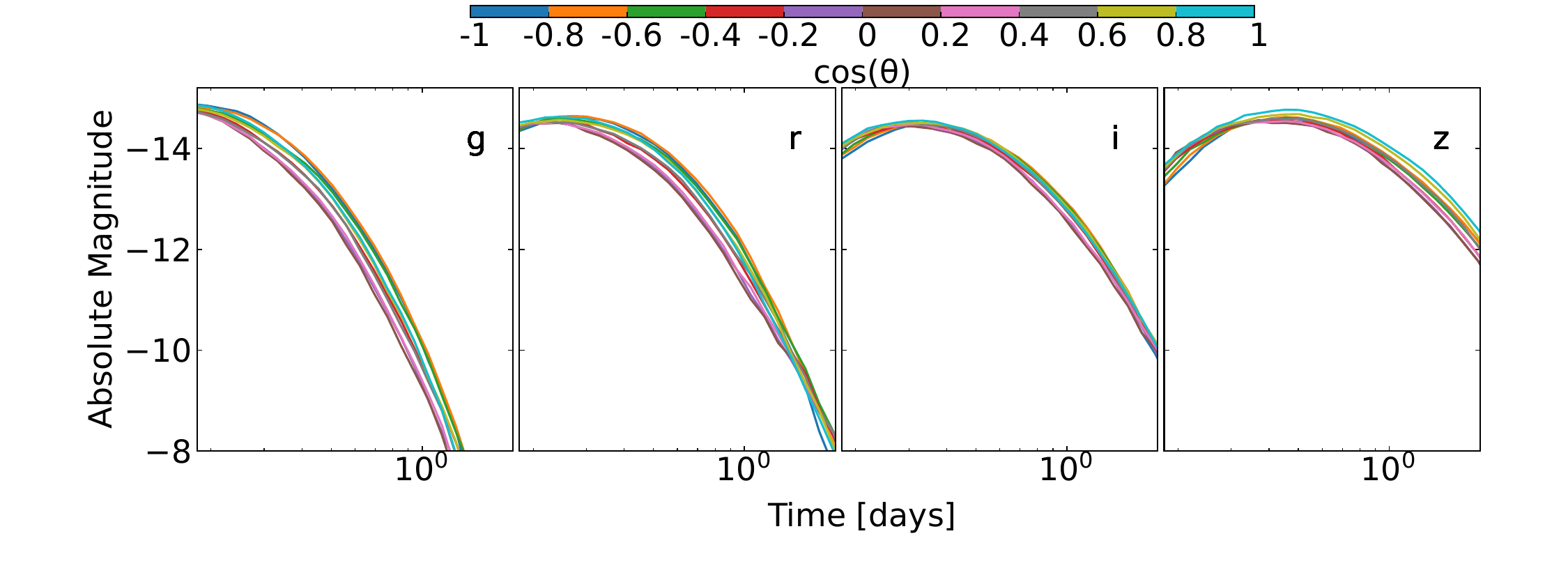}

\caption{Angle-dependent band-limited light curves.
}

\label{fig:bandlightcurves}
\end{figure*}

We show the $griz$ light curves for this simulation in Figure~\ref{fig:bandlightcurves}.
Similarly to the bolometric light curves, the band-limited light curves are not isotropic, but also do not show a very strong observer orientation dependence.
In all directions, the light curves initially peak in the bluer bands, and then become brighter in the redder bands,
showing similar behaviour to the observed blue to red colour evolution of AT2017gfo.
This was also found for this ejecta model in an approximate sense by \citet{collins2023a}.
Despite having a composition with a higher lanthanide fraction at the equator than at the poles,
we do not predict a significantly redder colour at the equator than the poles (see Figure~\ref{fig:colourevolution} {in Appendix~\ref{sec:supplementalfigs}}).
Although, we note that the atomic data considered in this simulation does not include actinides (see \citealt{shingles2023a} for details of the atomic data), and therefore the opacity may be underestimated.
This suggests that in a 3D simulation, a high lanthanide fraction at the equator does not necessarily lead to a significantly redder spectral energy distribution (SED) viewing from the equator than viewing from the poles.

\begin{figure*}
\centering
\subcaptionbox{Positive pole ($\mu_\mathrm{obs}$: $0.8 <$ cos($\mathrm{\theta}$) $< 1$)}
{\includegraphics[width=0.33\textwidth]{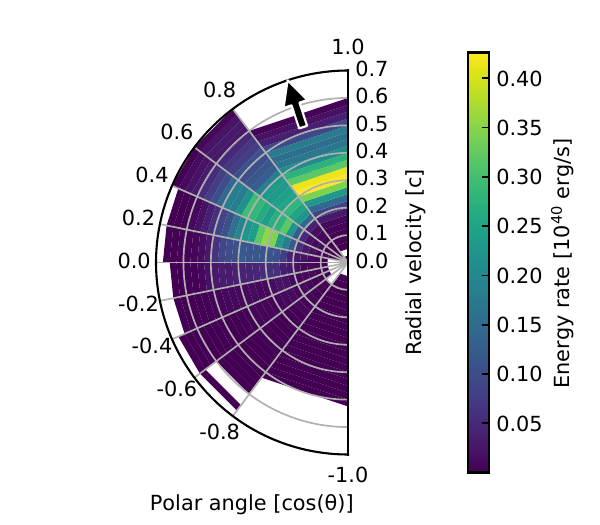}}
\subcaptionbox{Negative pole ($\mu_\mathrm{obs}$: $-1 <$ cos($\mathrm{\theta}$) $< -0.8$)}{\includegraphics[width=0.33\textwidth]{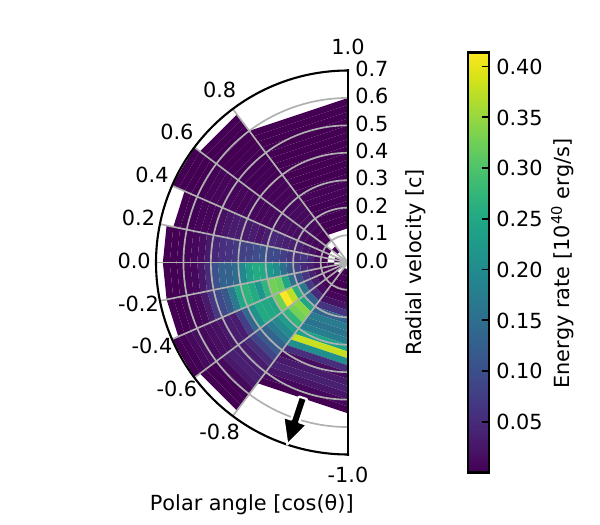}}
\subcaptionbox{Equator ($\mu_\mathrm{obs}$: $0 <$ cos($\mathrm{\theta}$) $< 0.2$)}{\includegraphics[width=0.33\textwidth]{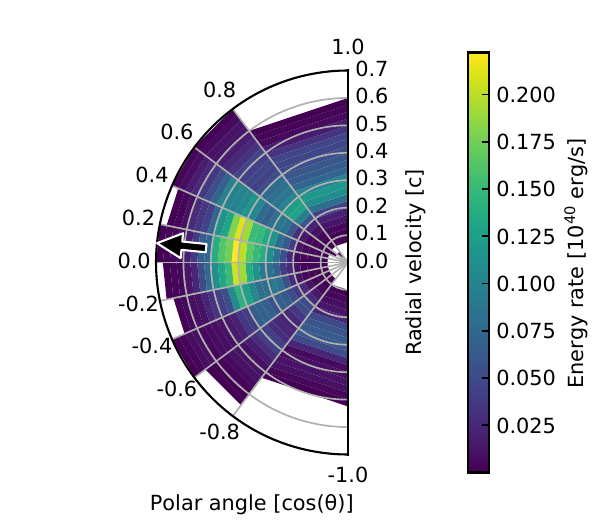}}

\caption{Radial velocity and polar angle of the location in the ejecta ($\mu_\mathrm{ej}$) where radiation last interacted ($v_\mathrm{i}$) before escaping towards an observer in the direction, $\mu_\mathrm{obs}$, (arriving at the observer at 0.4 days) viewing towards the poles (a and b) and viewing towards the equator (c). I.e., we construct a 2D histogram of the location and radial velocity where Monte Carlo packets of radiation escaping in a given direction last interacted with the ejecta, and weight each bin by the energy represented by the escaping packets.
The arrow indicates the direction, $\mu_\mathrm{obs}$, of the observer.
}

\label{fig:spherical_hist_allpkts}
\end{figure*}

The reason for this behaviour can be seen in Figure~\ref{fig:spherical_hist_allpkts}, where we show the location in the ejecta (in velocity space, assuming homologous expansion) where radiation last interacted before being emitted towards an observer viewing from a polar or an equatorial direction.
The radiation viewed from a given direction has been emitted from a broad range of ejecta,
both parallel and perpendicular to the line of sight.
Viewed from an equatorial direction, radiation is emitted from high opacity, lanthanide rich regions of the ejecta near the equator, but in addition to this, 
radiation is also emitted towards an observer at the equator from lower opacity regions (with lower lanthanide fractions) of the ejecta near the poles.
We therefore find that the asymmetry of the ejecta (in mass distribution and the variation in $Y_e$) does not strongly influence the anisotropy of the light curves for varying observer orientations,
since an observer does not view radiation emitted from only one region of the ejecta.

\subsubsection{Observer orientation variation in spectra}

\begin{figure}

\includegraphics[width=0.5\textwidth]{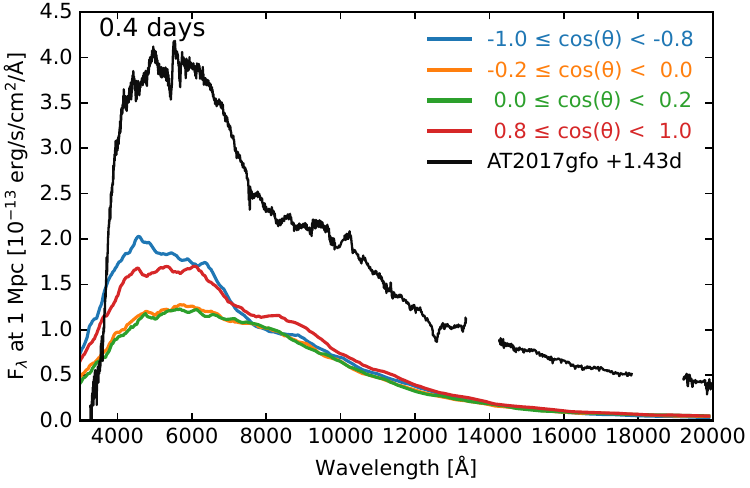}
\includegraphics[width=0.5\textwidth]{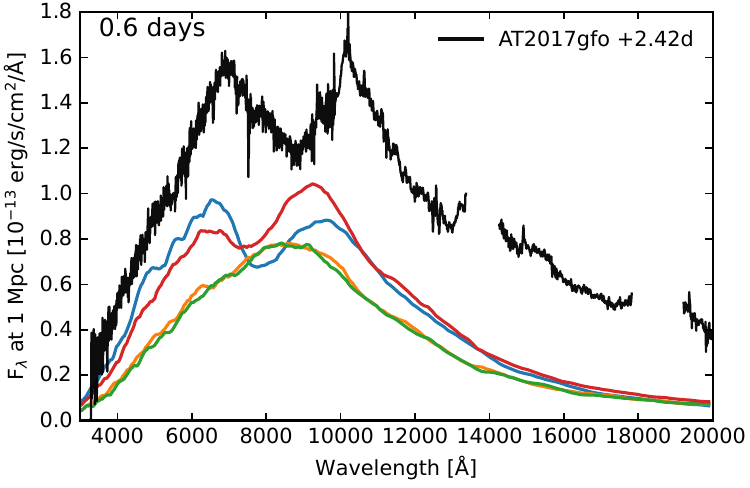}

\caption{Simulated spectra in polar and equatorial directions ($\mu_\mathrm{obs}$ is listed in the figure) at 0.4 days (upper) and 0.6 days (lower) compared to the spectra of AT2017gfo at 1.43 days and 2.42 days, respectively.}

\label{fig:spectralsequenceangles}
\end{figure}

As discussed by \citet{shingles2023a}, the model predicts that the spectra would not appear the same to an observer viewing towards the poles as to an observer near the equator.
We show the viewing-angle dependent spectra in Figure~\ref{fig:spectralsequenceangles}.
The model spectra show phases that resemble the observed spectra of AT2017gfo in the direction of the poles (see \citealt{shingles2023a}), however, at earlier times than those observed.
The more rapid evolution is likely due to the lower mass of our model compared to the inferred mass of AT2017gfo (see discussion by \citealt{shingles2023a}).
The spectra in the directions of the poles show a feature resembling a P-Cygni profile, which in the spectra of AT2017gfo has been suggested to be \ion{Sr}{II}.
As indicated by the band-limited light curves, the SEDs at the poles and equator peak at similar wavelengths,
however, at the equator the spectra are relatively featureless {(overlapping emission and absorption as well as Doppler broadening leads to no clear signatures of individual species, see \citealt{shingles2023a})} and fainter than the directions near the poles (Figure~\ref{fig:spectralsequenceangles}).
Therefore, our model spectra are significantly dependent on observer orientation.
We now  examine whether determination of the asymmetry parameter, $\Upsilon$, for the inferred photospheric velocities from these synthetic spectra can yield results consistent with observational constraints.

\subsection{Inferring photospheric velocities}
\label{sec:geometry-epm}

In this section we apply the same method as \citet{sneppen2023a} to our simulated spectra to determine the level of symmetry that would be inferred.
We use our model spectra at 0.4 days and 0.6 days, since these resemble the spectra of AT2017gfo at 1.43 days and 2.43 days, which were analysed by \citet{sneppen2023a}. We refer to the model spectrum at 0.4 days as epoch 1 and at 0.6 days as epoch 2.

\subsubsection{Photospheric velocity from P-Cygni feature}

\begin{figure*}

\includegraphics[width=0.95\textwidth]{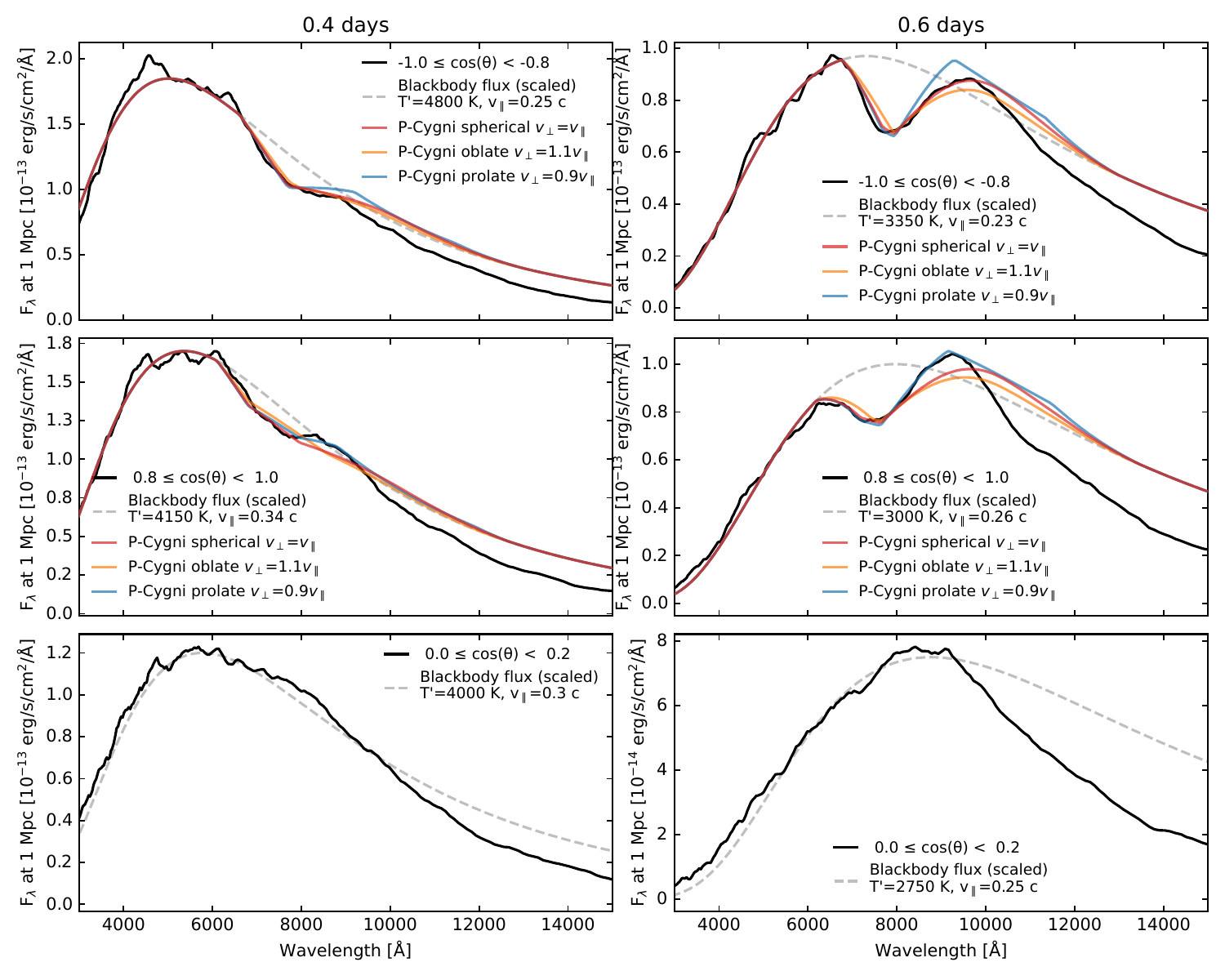}

\caption{Model spectra at epoch 1 (0.4 days; left panels) and epoch 2 (0.6 days; right panels) for polar (upper and middle panels) and equatorial (lower panel) observer directions ($\mu_\mathrm{obs}$, listed in each panel).
Dashed lines show the blackbody distribution (in the co-moving frame of the ejecta transformed into the rest frame of the observer, scaled to match the brightness of the synthetic spectra) best matching the spectra, where the ejecta temperature, T$^\prime$, and the parallel photospheric velocity, $v_\parallel$ are listed in each panel.
The blackbody distribution is modified to include a P-Cygni profile for \ion{Sr}{II}, considering spherical, prolate or oblate ejecta.
For the spectra at the equator, indicative photospheric velocities are chosen for the relativistic correction to the blackbody distribution, since photospheric velocity cannot be inferred from a spectral feature at the equator.
}

\label{fig:spec_BB_Pcygni}
\end{figure*}

The most prominent feature in the spectra of AT2017gfo has been suggested to be a \ion{Sr}{II} P-Cygni feature.
Our simulated spectra show a similar feature (Figure~\ref{fig:spectralsequenceangles}), as discussed by \citet{shingles2023a}.
We infer $v_\parallel$ from the simulated feature, assuming it can be modelled as a simple P-Cygni feature dominated by \ion{Sr}{II}, however,
we note that in the simulation the feature is actually a blend of features, predominantly \ion{Sr}{II}, \ion{Y}{II} and \ion{Zr}{II}.
Figure~\ref{fig:spec_BB_Pcygni} shows the \mbox{P-Cygni} profiles used to infer $v_\parallel$.
The P-Cygni profiles were generated using a line profile calculator\footnote{Available from \href{https://github.com/unoebauer/public-astro-tools}{https://github.com/unoebauer/public-astro-tools}.
For the \mbox{P-Cygni} profile considering an elliptical photosphere we use the version of this modified by \citet[][see their methods section]{sneppen2023a}, available from
\href{https://github.com/Sneppen/Kilonova-analysis}{https://github.com/Sneppen/Kilonova-analysis}.} based on the Elementary Supernova Model of \citet{jeffery1990a}.
We assume the \ion{Sr}{II} triplet lines to be of similar strength, and that one line (which we choose to be the mid-wavelength line at 10327.31 \AA) is representative of the triplet.
The P-Cygni profile is characterised by the rest wavelength, the optical depth and the velocity of the ejecta.
The velocities measured from the P-Cygni feature are listed in Table~\ref{tab:geometryparameters}.
Since the spectra in the equatorial lines of sight do not show any clear features, a velocity cannot be obtained in the same way from these synthetic spectra.

\begin{table*}
\centering
\caption{Quantities for inferring the geometry of our simulation for an observer viewing the simulation from a direction, $\mu_\mathrm{obs}$, where we give the range in polar-angle of $\mu_\mathrm{obs}$.
The parameters listed include the following.
Temperature in the co-moving frame of the ejecta, T$^\prime$, inferred from fitting a blackbody $B(\lambda, \mathrm{T}^\prime)$, transformed into the rest frame of the observer, to the simulated spectra (see Figure~\ref{fig:spec_BB_Pcygni}).
Inferred photospheric velocity, $v_\parallel$, from the P-Cygni feature, and the luminosity that would be inferred for these photospheric velocities using the expanding photosphere method ($L_\lambda^{\mathrm{BB}}$), which can be compared to the simulated luminosity $L^{{\mathrm{bol}}}$.
Perpendicular velocity, $v_\perp$, inferred from the simulated luminosity ($L^{\text{bol}}$) using the EPM.
Asymmetry index, $\Upsilon_{v,\mathrm{ph}}$, inferred from $v_{\|}$ and $v_{\perp}$.
Note that by 0.6 days the EPM does not predict reliable velocities for our simulation, which is the reason for the higher level of inferred asymmetry at this time.
}
\label{tab:geometryparameters}
\begin{tabular}{ccccccccc}
\hline
Epoch & Time  & $\mu_\mathrm{obs}$ & T$^\prime$ &  $v_\parallel$ (P-Cygni) & $v_\perp$ (EPM) & $\Upsilon_{v,\mathrm{ph}}$ & $L_\lambda^{\mathrm{BB}}$ (EPM) & $L^{\text{bol}}$    \\
 & {[d]}  &  [$\mathrm{cos(\theta)}$] & [K] & [c] &  [c] &  &  [erg s$^{-1}$] &  [erg s$^{-1}$]    \\
\hline
 1 & 0.4 & [$0.8, 1.0$] ($+$ve pole)   &  4150  & 0.34  & 0.33 & $-0.015$ & $1.54\times10^{41}$  & $1.43\times10^{41}$           \\
 1 & 0.4 & [$-1.0, -0.8$] ($-$ve pole) &  4800  & 0.25  & 0.27 &  $0.038$ & $1.26\times10^{41}$  & $1.46\times10^{41}$           \\
 1 & 0.4 & [$0.0, 0.2$] (eq.)        &  4000  &       & 0.32 &        &                      & $1.12\times10^{41}$           \\
 1 & 0.4 & [$-0.2, 0.0$] (eq.)       &  4000  &       & 0.32 &        &                      & $1.14\times10^{41}$           \\
\\
 2 & 0.6 & [$0.8, 1.0$] ($+$ve pole)   &  3000  &  0.26 & 0.38 & $0.19$   & $4.69\times10^{40}$  & $1.02\times10^{41}$           \\
 2 & 0.6 & [$-1.0, -0.8$] ($-$ve pole) &  3350  &  0.23 & 0.31 & $0.15$   & $5.42\times10^{40}$  & $9.90\times10^{40}$           \\
 2 & 0.6 & [$0.0, 0.2$] (eq.)        &  2750  &       & 0.40 &        &                      & $7.65\times10^{40}$           \\
 2 & 0.6 &  [$-0.2, 0.0$] (eq.)      &  2750  &       & 0.41 &        &                      & $7.80\times10^{40}$           \\
\hline    
\end{tabular}
\end{table*}

\subsubsection{Photospheric velocity from EPM}
\label{sec:inferred-perp-radius}

Following \citet{sneppen2023a},
we infer $v_\perp$ using the expanding photosphere method (EPM).
Assuming the emitting region to be a sphere of radius $R_{\mathrm{ph}}$,
emitting as a blackbody $B(\lambda, T^\prime)$,
at wavelength, $\lambda$, where $T^\prime$ is the inferred temperature in the co-moving frame of the ejecta, the inferred luminosity, $L_\lambda^{\mathrm{BB}}$, is given by
\begin{equation}
\label{eq:L_BB}
    L_\lambda^{\mathrm{BB}}=4 \pi R_{\mathrm{ph}}^2 \pi B(\lambda, T^\prime),
\end{equation}
where the blackbody flux emitted in the co-moving frame of the ejecta has been transformed into the rest frame of the observer \citep[see][]{sneppen2023a, sneppen2023c}.
The EPM assumes that $L_\lambda^{\mathrm{BB}}$ can be equated to the pseudo-bolometric luminosity,
\begin{equation}
\label{eq:L_bol}
    L^{\text{bol}}=4 \pi {D}_{\mathrm{L}}^2 F_\lambda,
\end{equation}
where $F_\lambda$ is the flux and $D_\mathrm{L}$ is the luminosity distance.
The EPM assumes homologous expansion, 
\begin{equation}
R_{\mathrm{ph}} = v_{\mathrm{ph}} t,
\end{equation}
where $ v_{\mathrm{ph}}$ is the velocity of the photosphere and $t$ is the time since explosion.

Using the EPM to infer $ v_{\mathrm{ph}}$, we obtain $v_\perp$ from the inferred blackbody temperature (by matching blackbody distributions to the synthetic spectra -- see Figure~\ref{fig:spec_BB_Pcygni}) and the luminosity obtained from the simulation.
The velocities inferred are listed in Table~\ref{tab:geometryparameters}.
We focus on fitting the blackbody distributions to the {UV and optical, rather than to the IR since the simulated
spectra appear to be well described by a blackbody at bluer wavelengths. However, compared to this blackbody fitting the spectra show a flux deficit in the IR}
(unlike AT2017gfo).
If a blackbody is fit to the IR flux, the peak of the blackbody is much too blue compared to the synthetic spectra.

At epoch 1, the values of $v_\perp$ obtained from the EPM
are similar to
the velocities, $v_\parallel$, inferred from the simulated spectral feature (e.g.,~0.34c compared to 0.33c at the positive pole).
At epoch 2, however, the velocities, $v_\perp$, inferred using the EPM are much higher than those inferred from the spectral feature, $v_\parallel$, (e.g., 0.38c compared to 0.26c at the positive pole).
Additionally, the inferred photospheric velocity from the EPM increases from epoch~1 to epoch~2,
in contradiction to the velocities from the spectral feature, the ejecta velocity radiation was emitted from (see Section~\ref{sec:comparelineveltosimulated}),
and the expectation that the photosphere would most likely recede with time.
At our simulated epoch~2, the spectrum is not well matched by a blackbody distribution
at wavelengths redder than $\sim 10000$~\AA \ (unlike AT2017gfo).
This likely explains why the EPM does not produce reasonable values of the
photospheric velocity for our simulation at epoch 2.
To an observer, it would be apparent that the spectra are not well represented by a blackbody at this time, and that the photospheric velocity inferred from
the EPM is not realistic.
\citet{dessart2005a} noted that the EPM is best used at early times when the spectrum is closest to a blackbody.
It is possible that the synthetic spectra show poorer agreement with a blackbody compared to AT2017gfo because the atomic data is not complete, for example, we do not include actinides.
Therefore opacity could be missing from the simulation.
Additionally, we only consider dynamical ejecta, which may also be responsible for the larger deviation from a blackbody.

\subsubsection{Geometry implied by $v_\|$ and $v_\perp$}
\label{sec:geometry_EPM_Pcygni}

Following \citet{sneppen2023a},
we use the asymmetry index,
\begin{equation}
    \Upsilon_{v,\mathrm{ph}}=\frac{v_{\perp}-v_{\|}}{v_{\perp}+v_{\|}},
\end{equation}
to quantify the degree of sphericity implied by our synthetic spectra,
which we show in Table~\ref{tab:geometryparameters}.
At epoch 1, a high level of symmetry is inferred from both polar directions.
Indeed, the inferred symmetry at the positive pole is within the uncertainty of the sphericity inferred by \citet{sneppen2023a}
for AT2017gfo ($\Upsilon=0.00 \pm 0.02$).
This demonstrates that aspherical ejecta can lead to directions that can appear near-symmetrical to an observer, as quantified by the $\Upsilon_{v,\mathrm{ph}}$ measurement.

At epoch 2, however, the inferred values of $\Upsilon_{v,\mathrm{ph}}$ are much higher.
As discussed in Section~\ref{sec:inferred-perp-radius}, this is likely because the simulated spectra are no longer well matched by a blackbody and the inferred perpendicular velocities are much higher than expected. 
This was not the case for AT2017gfo, during the epochs analysed by \citet{sneppen2023a}, where the spectra continued to resemble a blackbody until later times than in our simulation.

\subsubsection{Distance estimate}

The EPM can be used to infer the luminosity distance, D$_\mathrm{L}$, by equating
Equations~\ref{eq:L_BB} and \ref{eq:L_bol}.
We test how accurately the distance can be inferred for our simulated spectra.
We set the distance to the simulated spectra as 1 Mpc, and the inferred distances using the EPM are listed in Table~\ref{tab:distance_estimate}.
For this calculation, we assume a spherical photosphere ($v_\parallel = v_\perp$) and use the velocity, $v_\parallel$, inferred from the P-Cygni profile (Table~\ref{tab:geometryparameters}) to determine R$_\mathrm{ph}$.
We use the temperatures inferred from the blackbody distributions matching the simulated spectra (Table~\ref{tab:geometryparameters}), and the flux from the simulated spectra.

At epoch 1, the distance can be inferred to a good degree of accuracy (within 4--7 per cent),
however, at epoch 2 when the spectra are no longer well matched by a blackbody the distance estimate is much more uncertain (>25 per cent error) and underestimated.
For each epoch, the direction with the more spherical value of $\Upsilon_{v,\mathrm{ph}}$ (Table~\ref{tab:geometryparameters}) gives a closer estimate of distance to the actual distance, possibly indicating that testing the sphericity implied by $v_\parallel$ and $v_\perp$ could be a test for how good a distance estimate can be obtained, but this should be investigated for more models in future.

Both the P-Cygni profile analysis and the EPM assume that the photosphere is a sharp boundary at a single velocity.
We discuss in Section~\ref{sec:comparelineveltosimulated} that in our simulation radiation escapes from a broad range of ejecta velocities,
which is likely also the case for AT2017gfo, indicating that the photosphere is not a sharp boundary.
However, even with this simple assumption an accurate distance estimate can be obtained while the spectra resemble a blackbody.

\begin{table}
\centering
\caption{Luminosity distance, D$_\mathrm{L}$, estimated using the EPM in the direction $\mu_\mathrm{obs}$. We set the distance to our simulated spectra as 1 Mpc.
The photospheric velocity considered is that inferred from the P-Cygni profile (Table~\ref{tab:geometryparameters}).
}
\label{tab:distance_estimate}
\begin{tabular}{cccc}
\hline
Epoch & Time  & $\mu_\mathrm{obs}$  & D$_\mathrm{L}$   \\
      & {[d]}  &  [$\mathrm{cos(\theta)}$]  & [Mpc] \\
\hline
 1 & 0.4 & [$0.8, 1.0$] ($+$ve pole) &   1.04 \\
 1 & 0.4 & [$-1.0, -0.8$] ($-$ve pole)  &   0.93       \\
 \\
 2 & 0.6 & [$0.8, 1.0$] ($+$ve pole)  &  0.68 \\
 2 & 0.6 & [$-1.0, -0.8$] ($-$ve pole) &  0.74    \\

\hline    
\end{tabular}
\end{table}

\subsection{Isotropy of spectrum forming region}
\label{sec:comparelineveltosimulated}

\begin{table*}
\centering
\caption{Mean ejecta radial velocities of the last interaction underwent by a Monte Carlo packet of radiation ($\bar{v}_\mathrm{i}$) before escaping towards an observer in the direction, $\mu_\mathrm{obs}$.
$\bar{v}_\mathrm{i}$ gives an indication of the spectrum-forming region in the ejecta.
We show this for packets of radiation escaping at all wavelengths and for packets of radiation last absorbed by the \ion{Sr}{II} triplet ($\bar{v}_\mathrm{i}^\mathrm{Sr}$) before being re-emitted and escaping towards an observer. 
Also listed is the standard deviation $\sigma$ in $v_\mathrm{i}$ and $v_\mathrm{i}^\mathrm{Sr}$.
}
\label{tab:measuredvelocities}
\begin{tabular}{ccccccc}
\hline
Epoch & Time & $\mu_\mathrm{obs}$  &  $\bar{v}_\mathrm{i}$ &  $\sigma_{v\mathrm{,i}}$ & $\bar{v}_{\mathrm{i}}^\mathrm{Sr}$ & $\sigma_{v\mathrm{,i}}^\mathrm{Sr}$  \\ 
 & {[d]} & [$\mathrm{cos(\theta)}$]       & (all wavelengths) [c] & [c] & (\ion{Sr}{II} triplet absorption) [c] & [c]\\ 
\hline
1 & 0.4 & [$0.8, 1.0$] ($+$ve pole) & 0.35  &   0.10 & 0.42  &    0.09 \\ 
1 & 0.4 & [$-1.0, -0.8$] ($-$ve pole)  & 0.32 &  0.09  & 0.37  &   0.08   \\ 
1 & 0.4 & [$0.0, 0.2$] (eq.) & 0.37  & 0.10   & 0.44  &   0.08  \\ 
1 & 0.4 & [$-0.2, 0.0$] (eq.) & 0.37  &  0.10  & 0.43 &   0.08    \\ 
\\
2 & 0.6 & [$0.8, 1.0$] ($+$ve pole) & 0.33  &  0.10  & 0.37  &   0.08  \\
2 & 0.6 & [$-1.0, -0.8$] ($-$ve pole) & 0.30 &  0.09  & 0.33   &   0.08   \\ 
2 & 0.6 & [$0.0, 0.2$] (eq.) & 0.34 &  0.10  & 0.38  &  0.08    \\ 
2 & 0.6 & [$-0.2, 0.0$] (eq.) & 0.34 &  0.10  & 0.37  &   0.09    \\ 
\hline

\end{tabular}
\end{table*}

\begin{table*}
\centering
\caption{Asymmetry index (Equation~\ref{eq:asymparam-lastinteraction}) for the mean last-interaction velocities ($\bar{v}_\mathrm{i}$), listed in Table~\ref{tab:measuredvelocities}, of packets escaping
into the solid-angle $\mu_\mathrm{obs}$, in a polar orientation ($\bar{v}_{\mathrm{i,pole}}$) and an equatorial orientation ($\bar{v}_{\mathrm{i,eq}}$).
This is shown for packets of radiation escaping at all wavelengths, and for radiation that was last absorbed by the \ion{Sr}{II} triplet.
}
\label{tab:asymparam-lastinteraction_vel}
\begin{tabular}{cccccc}
\hline
 Epoch & Time  & $\mu_\mathrm{obs}$ (pole)  & $\mu_\mathrm{obs}$ (equator) & $\Upsilon_{\bar{v},\mathrm{i}}$  & $\Upsilon_{\bar{v},\mathrm{i}}^\mathrm{Sr}$ \\
 & {[d]}  &  [$\mathrm{cos(\theta)}$]  &  [$\mathrm{cos(\theta)}$]  & (All wavelengths) & (\ion{Sr}{II} triplet absorption) \\
\hline
 1 & 0.4 & [$0.8, 1.0$] ($+$ve pole) & [$0.0, 0.2$] (eq.)  &  0.028 & 0.023  \\
 1 & 0.4 & [$-1.0, -0.8$] ($-$ve pole)  & [$-0.2, 0.0$] (eq.)  &   0.072 & 0.075       \\

 \\
 2 & 0.6 & [$0.8, 1.0$] ($+$ve pole)  &   [$0.0, 0.2$] (eq.) & 0.015 & 0.013    \\
 2 & 0.6 & [$-1.0, -0.8$] ($-$ve pole)  &  [$-0.2, 0.0$] (eq.) &  0.063 & 0.057  \\

\hline    
\end{tabular}
\end{table*}

\subsubsection{Last-interaction velocities}

\begin{figure}

\subcaptionbox{All wavelengths}{\includegraphics[width=0.48\textwidth]{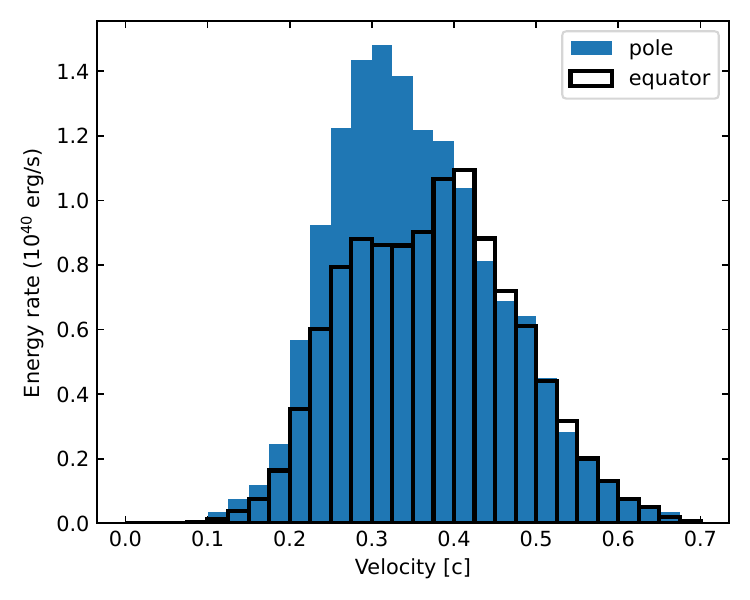}}
\subcaptionbox{\ion{Sr}{II} triplet absorption}{\includegraphics[width=0.48\textwidth]{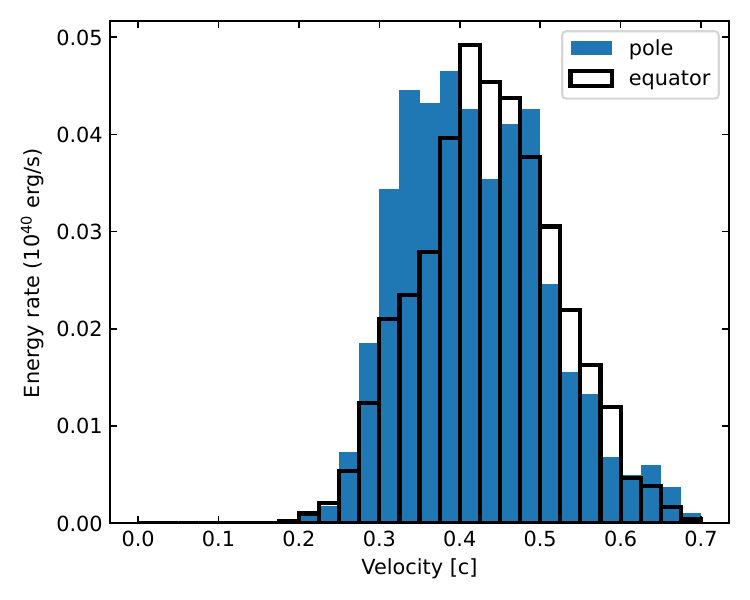}}

\caption{Histograms showing the last-interaction velocities, (a) $v_\mathrm{i}$ and (b)~$v_\mathrm{i}^\mathrm{Sr}$, weighted by the energy represented by the {escaped} packets of radiation, {arriving at} an observer at the pole ($\mu_\mathrm{obs}$: $0.8 < \mathrm{cos}(\theta) < 1.0$) or equator ($\mu_\mathrm{obs}$:~$0 < \mathrm{cos}(\theta) < 0.2$) at 0.4 days.
}

\label{fig:hist-lineformingregion}
\end{figure}

To give an indication of how well the asymmetry parameter, $\Upsilon_{v,\mathrm{ph}}$, from the inferred photospheric velocities 
represents the symmetry of radiation leaving the simulation,
we extract from the \textsc{artis} calculation the ejecta radial velocity~($v_\mathrm{i}$) at which each Monte Carlo packet last interacted with the ejecta before escaping towards an observer.
\textsc{artis} does not impose any photospheric boundary condition on the simulation,
but the distribution of $v_\mathrm{i}$ provides an indication of where radiation-matter interactions are occurring
and thus the location of the spectrum forming region (see Figure~\ref{fig:spherical_hist_allpkts}).
The mean, energy weighted radial velocity at which radiation packets last interacted with the ejecta ($\bar{v}_\mathrm{i}$) is listed in Table~\ref{tab:measuredvelocities} for radiation {that has escaped the ejecta, travelling} towards an observer at a polar or equatorial orientation (i.e.,~packets of radiation {that have escaped} into the given solid-angle bin, $\mu_\mathrm{obs}$) at epochs 1 and 2.
The range of $v_\mathrm{i}$ of packets {that have escaped} in a given direction can be seen in Figures~\ref{fig:spherical_hist_allpkts} and \ref{fig:hist-lineformingregion} for epoch 1, giving an indication of the extent of the spectrum forming region.
{In particular, at the equator the radiation escapes from an extremely broad range of velocities, as can be seen by the more flat topped distribution of $v_\mathrm{i}$ in Figure~\ref{fig:hist-lineformingregion} at the equator, compared to at the pole which is more sharply peaked around 0.3c.}

\begin{figure*}

\subcaptionbox{Positive pole ($\mu_\mathrm{obs}$: $0.8 <$ cos($\mathrm{\theta}$) $< 1$)}{\includegraphics[width=0.33\textwidth]{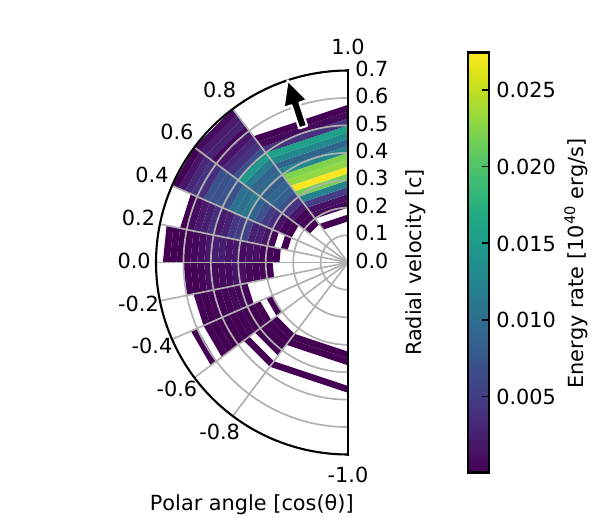}}
\subcaptionbox{Negative pole ($\mu_\mathrm{obs}$: $-1 <$ cos($\mathrm{\theta}$) $< -0.8$)}{\includegraphics[width=0.33\textwidth]{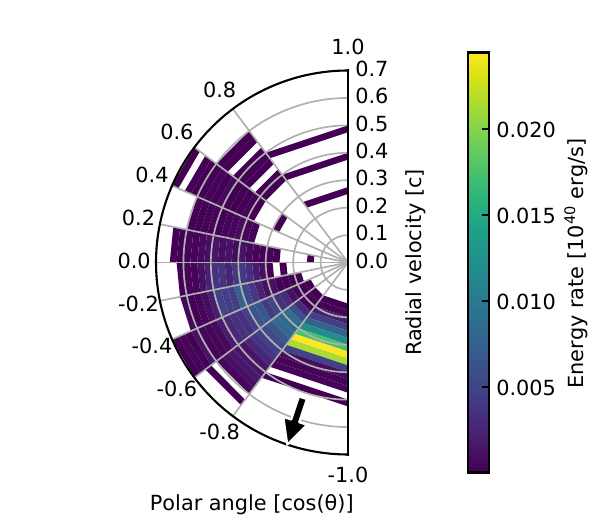}}
\subcaptionbox{Equator ($\mu_\mathrm{obs}$: $0 <$ cos($\mathrm{\theta}$) $< 0.2$)}{\includegraphics[width=0.33\textwidth]{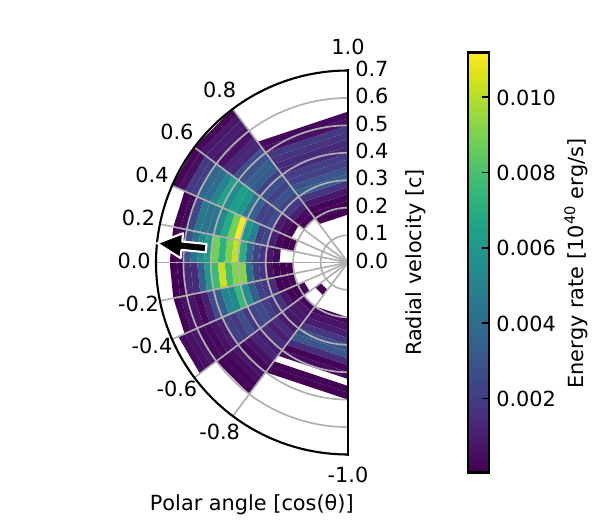}}

\caption{
Radial velocity and polar angle of the location in the ejecta ($\mu_\mathrm{ej}$) where radiation was last absorbed by the \ion{Sr}{II} triplet ($v_\mathrm{i}^\mathrm{Sr}$) before being re-emitted towards an observer in the direction, $\mu_\mathrm{obs}$, (arriving at the observer at 0.4 days) viewing towards the poles (a and b) and towards the equator (c).
The arrow indicates the direction, $\mu_\mathrm{obs}$, of the observer.
}

\label{fig:absorptionvelSrII}
\end{figure*}

We also show in Figures~\ref{fig:hist-lineformingregion} and \ref{fig:absorptionvelSrII} the range of ejecta velocities where the last absorption process underwent by a packet was with the \ion{Sr}{II} triplet (${v}_{\mathrm{i}}^\mathrm{Sr}$), with wavelengths of 10036.65, 10327.31 and 10914.87~\AA,
immediately before being re-emitted towards an observer in a given direction.
The mean, energy weighted radial velocities where radiation was last absorbed by the \ion{Sr}{II} triplet, $\bar{v}_{\mathrm{i}}^\mathrm{Sr}$,
are listed in Table~\ref{tab:measuredvelocities}.

As discussed by \citet{shingles2023a},
\ion{Sr}{II} is not the only species responsible for shaping the predicted feature. 
The \ion{Sr}{II} triplet absorption is, however, at the velocities required to match the feature. 
The values of $\bar{v}_{\mathrm{i}}^\mathrm{Sr}$ are higher than the values of $\bar{v}_{\mathrm{i}}$ (e.g., 0.42c compared to 0.35c, see Table~\ref{tab:measuredvelocities}).
This is broadly consistent with the general principles of the simple P-Cygni model adopted to fit this feature, i.e., photon interactions in the Sr line are generally occurring in a spatially extended line-forming region that extends above the region in which the pseudo-continuum forms.

As expected, the values of $\bar{v}_{\mathrm{i}}$ and $\bar{v}_{\mathrm{i}}^\mathrm{Sr}$ decrease from epoch 1~to~2, indicating that the spectrum-forming region recedes from epoch 1 to 2 in the simulation, verifying that the velocities inferred at epoch 2 using the EPM are too high. 

\subsubsection{Symmetry of last-interaction velocities}

To quantify the level of symmetry shown by the mean last-interaction velocities, we compare $\bar{v}_\mathrm{i}$ of radiation that escaped towards an observer at a polar orientation ($\bar{v}_{\mathrm{i,pole}}$) to $\bar{v}_\mathrm{i}$ of radiation escaping towards an observer at an equatorial orientation ($\bar{v}_{\mathrm{i,eq}}$), using
\begin{equation}
\label{eq:asymparam-lastinteraction}
    \Upsilon_{\bar{v},\mathrm{i}}=\frac{\bar{v}_{\mathrm{i,eq}}-\bar{v}_{\mathrm{i,pole}}}{\bar{v}_{\mathrm{i,eq}}+\bar{v}_{\mathrm{i,pole}}},
\end{equation}
which is listed in Table~\ref{tab:asymparam-lastinteraction_vel}.

At both epochs, $\Upsilon_{\bar{v},\mathrm{i}}$
indicates a high degree of symmetry, with relatively low values for $\Upsilon_{\bar{v},\mathrm{i}}$.
However, $\Upsilon_{\bar{v},\mathrm{i}}$ indicates a slightly lower level of symmetry than determined in Section~\ref{sec:geometry_EPM_Pcygni} from the inferred photospheric velocities, $\Upsilon_{v,\mathrm{ph}}$, (at epoch 1), where $\Upsilon_{\bar{v},\mathrm{i}} > \Upsilon_{v,\mathrm{ph}}$.
$\Upsilon_{\bar{v},\mathrm{i}}$ indicates that the level of symmetry increases from epoch 1 to 2 in this model.

In agreement with $\Upsilon_{v,\mathrm{ph}}$ inferred from spectra at either pole at epoch 1,
a higher degree of symmetry (as indicated by $\Upsilon_{\bar{v},\mathrm{i}}$) is found
at the positive pole \mbox{($0.8 <$ cos($\mathrm{\theta}$) $< 1$)}, than at the negative pole \mbox{($-1 <$ cos($\mathrm{\theta}$) $< -0.8$)}.
The level of symmetry inferred from radiation escaping at all wavelengths $\Upsilon_{\bar{v},\mathrm{i}}$, is similar to that shown by only radiation that was last absorbed by the \ion{Sr}{II} triplet, $\Upsilon_{\bar{v},\mathrm{i}}^\mathrm{Sr}$.
Even though the distribution of Sr in the ejecta is highly asymmetrical (as quantified by $\Upsilon_\mathrm{M}^\mathrm{Sr}$ and $\Upsilon_{\mathrm{M,}v}^\mathrm{Sr}$), $\Upsilon_{\bar{v},\mathrm{i}}^\mathrm{Sr}$ is highly spherical.

The high level of symmetry shown by the mean last-interaction velocities, as quantified by $\Upsilon_{\bar{v},\mathrm{i}}$,
supports the high level of symmetry inferred from $v_\parallel$ and $v_\perp$ at epoch 1, measured from the P-Cygni feature and EPM. {$\Upsilon_{\bar{v},\mathrm{i}}$ remains highly spherical at epoch 2, showing that the simulation still exhibits a high level of symmetry beyond epoch 1. This implies that if the EPM was still valid at epoch 2 to infer $v_\perp$ then a high degree of sphericity could be found for this simulation.}.
This demonstrates that even when the mass distribution and elemental abundances in the ejecta are asymmetric, the line forming region can be relatively close to spherical.

\subsection{P-Cygni profile analysis}

\subsubsection{Inferring geometry from P-Cygni profile}

Following \citet{sneppen2023a}, we analyse the
shape of the \mbox{P-Cygni} profile in our synthetic spectra
to explore how well it can be used to constrain the geometry of our simulation.
Using the simple P-Cygni model,
we show the comparison of P-Cygni features with spherical and elliptical photospheres\footnote{See \citealt{sneppen2023a} for details of the modifications to the line profile calculator to represent an elliptical photosphere.
Note that we did not apply the MCMC fitting used by \citet{sneppen2023a} to fit the P-Cygni profile, but rather varied parameters manually to match the P-Cygni feature.} to our simulated spectra in Figure~\ref{fig:spec_BB_Pcygni}. 
We focus on matching the wavelengths of the absorption component of the P-Cygni profile and the blue side of the emission component, given that redward of the emission feature the spectra
are not well matched by the blackbody distribution (which was not the case for AT2017gfo).
We also note that the red side of the emission component is strongly blended with \ion{Ce}{III} in our simulation (see \citealt{shingles2023a}, figure 4).
Across both epochs, the feature in the simulated spectra at the negative pole is best matched by a P-Cygni feature with a spherical photosphere, and the spectral feature at the positive pole is best matched by a P-Cygni feature with a prolate photosphere.
This is surprising since the positive pole was inferred to show a higher degree of symmetry in Section~\ref{sec:geometry_EPM_Pcygni} (Table~\ref{tab:geometryparameters}).
Since opposite poles can not be fit by assuming the same geometry for the photosphere, this demonstrates that assuming a single photospheric geometry for the entire model is likely too simple an approximation.
The range of $v_\mathrm{i}$ (Figure~\ref{fig:hist-lineformingregion}) shows that radiation is emitted from a broad distribution of ejecta, which may not be well captured by assuming the photosphere can be modelled by a single velocity. 
We note, however, that it was possible for AT2017gfo to obtain a better fit with a P-Cygni profile than for our synthetic spectra (see \citealt{sneppen2023a}) and this analysis should be tested for models more closely resembling a blackbody in future.

The distributions of Sr, Y and Zr, which are predominantly responsible for shaping the simulated spectral feature, do not show a spherical distribution since significantly lower masses of these elements are ejected at the equator (Figure~\ref{fig:mass-histograms-elements}).
Since there is a direction for our asymmetric ejecta model at which the spectral feature is best matched by assuming a spherical photosphere,
this demonstrates that the underlying ejecta does not necessarily have to be symmetrical for the P-Cygni profile shape to appear consistent with a spherical model.

\subsubsection{P-Cygni profile shape for spherically symmetric ejecta}

\begin{figure}

\includegraphics[width=0.48\textwidth]{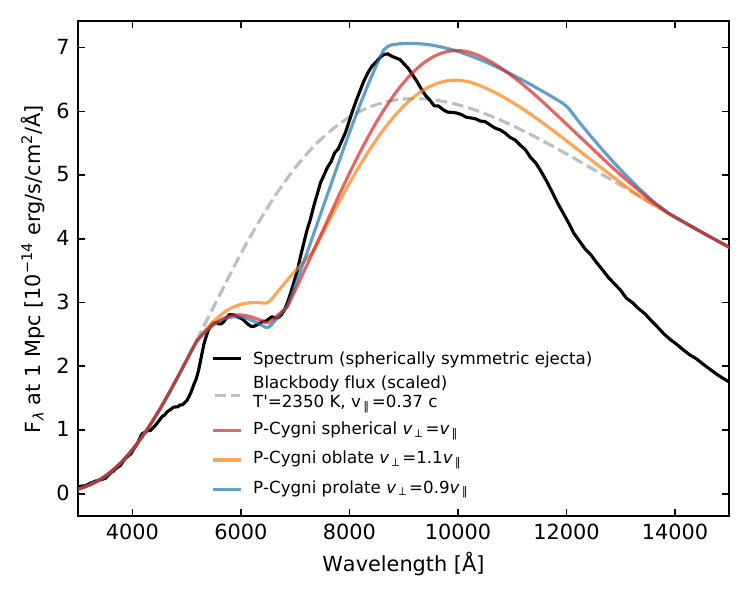}

\caption{Spectrum at 0.4 days from spherically symmetric ejecta, compared to P-Cygni profiles 
(assuming the feature can be fit by a simple \ion{Sr}{II} P-Cygni feature)
with a spherical, oblate or prolate photosphere.
The dashed line shows a blackbody distribution.
To plot the P-Cygni profiles, we modify the blackbody with the calculated line profile.
However, we note that in the 1D spherical simulation, the spectra are not well represented by a blackbody.}

\label{fig:1D-pcygni}
\end{figure}

Having shown that an aspherical model can produce line profiles that appear consistent with spherical ejecta,
we now explore whether the line shape predicted from a model with spherical ejecta is also well matched if fit with a simple P-Cygni model.
\citet{shingles2023a} present a model where they enforce spherically symmetric ejecta by constructing a 1D,
spherically averaged version of the 3D model we consider in this paper (1D AD2 from \citealt{shingles2023a}).
Both the mass distribution and elemental abundances are spherically averaged.
Imposing spherical symmetry on this model results in spectra and light curves that do not resemble those predicted for any observer direction from the full 3D simulation.
The 3D structures in this model are important in shaping the spectra. Only when the 3D structure is included is the model able
to produce synthetic observables that resemble AT2017gfo for this model.
We note, however, that 1D simulations can produce spectra resembling AT2017gfo (e.g., \citealt{watson2019a, gillanders2022a}).

As with our 3D model, we fit the spectral feature produced by the 1D spherically symmetric simulation at $\sim 6000$ -- $8000$ \AA \ with P-Cygni profiles assuming a spherical, prolate or oblate photosphere (Figure~\ref{fig:1D-pcygni}).
Note that in this 1D simulation, the `photosphere' (and spectrum-forming region) must be spherical.
The best match is with the P-Cygni profile assuming a prolate photosphere. The emission component (on the blue side) increases too steeply to be matched by a spherical or oblate photosphere.
However, given that the continuum is not
well described by a blackbody, the exact contribution of the P-Cygni profile is more uncertain and less constrained than in the 3D case.
Since we know that this simulation is spherical,
this indicates that the geometry of the photosphere assumed to calculate a simple P-Cygni profile may not be
a good test for spherically symmetric ejecta.

In both the 1D spherical simulation and the 3D asymmetric simulation, the spectral feature is actually a blend of features (primarily \ion{Sr}{II}, \ion{Y}{II}, \ion{Zr}{II} and \ion{Ce}{III} -- see \citealt{shingles2023a}), and the simple \ion{Sr}{II} P-Cygni calculation here is likely unable to capture the complex interplay of these spectral features.

\section{Conclusions}

We have analysed a 3D radiative transfer simulation of a kilonova carried out by
\citet{shingles2023a} to determine whether the simulation is compatible with the inferred symmetry constraints suggested for AT2017gfo by \citet{sneppen2023a}.

We have shown that although the ejecta from the neutron star merger model have moderate asymmetries in the mass ejected per solid-angle (e.g., $\Upsilon_\mathrm{M}$=0.33), the synthetic light curves
produced show a lower level of asymmetry (e.g. $\Upsilon_{\mathrm{bol}}=-0.12$ at 0.4 days)
than the level of asymmetry in the mass-distribution of the ejecta.
The mass ejected within the velocity range that is approximately within the line forming region for the first day of the kilonova shows a high degree of symmetry (e.g., $\Upsilon_{\mathrm{M},v}=0.037$ at the negative pole), which may lead to a higher level of symmetry in the synthetic observables than suggested by $\Upsilon_{\mathrm{M}}$,
however, the distribution of elements synthesised in the ejecta in this velocity region is very asymmetric (e.g., \mbox{$\Upsilon_{\mathrm{M},v}^\mathrm{Sr} = -0.71$} at the positive pole).
Radiation observed in a given line of sight is emitted from a broad range of regions in the ejecta, which has the effect of decreasing the observed anisotropy.
For example, the \ion{Sr}{II} triplet absorption is highly spherical (e.g., $\Upsilon_{v\mathrm{,i}}^\mathrm{Sr} = 0.023$ at the positive pole) despite the asymmetric distribution of Sr in the ejecta.
However, the spectra are not predicted to appear the same at the poles as at the equator. 
The equatorial spectra are relatively featureless in comparison to the spectra produced near the poles (as discussed by \citealt{shingles2023a}).

Following \citet{sneppen2023a},
we quantify the level of symmetry that would be inferred from our simulation from the photospheric velocities, $v_\parallel$ and $v_\perp$, obtained via simple fitting of the synthetic spectra (i.e., adopting similar methods as can be readily applied to real observations).
At the first epoch considered (0.4 days), the values inferred for $v_\perp$ are similar to those obtained for $v_\parallel$, indicating a high degree of sphericity, as quantified by $\Upsilon_{v,\mathrm{ph}}$.
From one pole, $\Upsilon_{v,\mathrm{ph}}$ is within the uncertainty inferred for AT2017gfo by \citet{sneppen2023a}.
This demonstrates that the synthetic observables can appear consistent with spherical ejecta when viewed from certain directions, even when the ejecta are asymmetric.

At the second epoch (0.6 days) the synthetic spectra are no longer well represented by a blackbody, and the inferred values of $v_\perp$ are too high.
However, this is not what was found by \citet{sneppen2023a} for the observations of AT2017gfo, where the spectra continue to resemble a blackbody until later times and the inferred photospheric velocities from AT2017gfo using the EPM were similar to
those obtained from the \ion{Sr}{II} P-Cygni feature over the two epochs considered by \citet{sneppen2023a}.

The high degree of symmetry determined from the inferred photospheric velocities, quantified by $\Upsilon_{v,\mathrm{ph}}$, at epoch 1 is supported by the level of symmetry in the mean last-interaction velocities extracted from the simulation, quantified by $\Upsilon_{\bar{v},\mathrm{i}}$.
The level of symmetry indicated by $\Upsilon_{\bar{v},\mathrm{i}}$ increases from epoch 1 to 2.

At epoch 1, while the spectra are well represented by a blackbody, the EPM can be used to infer the distance to our synthetic spectra to a good degree of accuracy (4--7 per cent).
At epoch 2, however, the distance is underestimated (>25 per cent).
In our simulation, where a higher degree of sphericity, $\Upsilon_{v,\mathrm{ph}}$, was inferred from the synthetic spectra, a more accurate estimate of the distance D$_\mathrm{L}$ was obtained.
This may suggest that $\Upsilon_{v,\mathrm{ph}}$ could be used as a test for how accurate a distance estimate can be obtained from the EPM, however, this should be investigated with future simulations.

We compare simple P-Cygni profile models, assuming the photosphere to be spherical or elliptical, to our simulated spectral feature.
The shape of the spectral feature produced at one pole
is best matched by a P-Cygni profile assuming a spherical photosphere.
However, from the opposite pole the shape was best matched by a \mbox{P-Cygni} model assuming a prolate photosphere.
Additionally, we found that for a spectrum from a simulation with spherically symmetric ejecta, the feature was best matched by a P-Cygni profile assuming a prolate photosphere.
These simulations suggest that fitting the profile shape alone may not be a robust test of spherical symmetry, although we note that the P-Cygni fits to our simulations are less certain than for AT2017gfo, due to the poorer match to a blackbody redward of the P-Cygni profile.

In our 3D simulation, there are lines of sight for which it could be inferred that the kilonova is highly spherical using the methods suggested by \citet{sneppen2023a}.
The level of symmetry of the last-interaction velocities ($\Upsilon_{v\mathrm{,i}}$) of radiation escaping from our simulation is highly spherical, despite the asymmetric ejecta.
This indicates that the combination of $v_\parallel$ and $v_\perp$ can indicate sphericity of the escaping radiation, however, this should not be interpreted to mean that the ejecta are necessarily spherically symmetric.
Not all directions in the simulation would appear as spherical as inferred for AT2017gfo.
If our simulation is representative of a real kilonova event, then we would expect that a high level of symmetry would not be inferred for all future observations (e.g., at the equator, where we do not predict a spectral feature from which to infer $v_\parallel$).
However, if all future observations appear as spherical as AT2017gfo then this could suggest that this simulations is too anisotropic.
More observations and simulations are required to understand the geometry of kilonovae.


\section*{Acknowledgements}
CEC, AB, OJ, GLi and VV acknowledge support by the European Research Council (ERC) under the European Union’s Horizon 2020 research and innovation program under grant agreement No. 759253.
CEC, AB, SAS, AS and DW acknowledge support by the European Research Council (ERC) under the European Union’s research and innovation program (ERC Grant HEAVYMETAL No. 101071865).
LJS, GLi, AF, GLe, GMP, and ZX acknowledge support by the European Research Council (ERC) under the European Union’s Horizon 2020 research and innovation program (ERC Advanced Grant KILONOVA No. 885281).
AB, CEC, GLi, AF, OJ, GLe, GMP, LJS, and ZX acknowledge support by Deutsche Forschungsgemeinschaft (DFG, German Research Foundation) - Project-ID 279384907 - SFB 1245 and MA 4248/3-1.
AB, GLi and VV acknowledge support by DFG - Project-ID 138713538 - SFB 881 (“The Milky Way System”, subproject A10).
AB, CEC, TS, AF, GLe, GMP, OJ, LJS, and ZX acknowledge support by the State of Hesse within the Cluster Project ELEMENTS.
The work of SAS was supported by the Science and Technology Facilities Council [grant numbers ST/P000312/1, ST/T000198/1, ST/X00094X/1].
The Cosmic Dawn
Center is funded by the Danish National Research Foundation under grant number 140.
The authors gratefully acknowledge the Gauss Centre for Supercomputing e.V.
(www.gauss-centre.eu) for funding this project by providing computing time
through the John von Neumann Institute for Computing (NIC) on the GCS
Supercomputer JUWELS at J\"ulich Supercomputing Centre (JSC).
This work was performed using the Cambridge Service for Data Driven Discovery (CSD3), part of which is operated by the University of Cambridge Research Computing on behalf of the STFC DiRAC HPC Facility (www.dirac.ac.uk). The DiRAC component of CSD3 was funded by BEIS capital funding via STFC capital grants ST/P002307/1 and ST/R002452/1 and STFC operations grant ST/R00689X/1. DiRAC is part of the National e-Infrastructure.
CEC and LJS are grateful for computational support by the VIRGO cluster at GSI.
NumPy and SciPy
\citep{oliphant2007a}, Matplotlib
\citep{hunter2007a} and \href{https://github.com/artis-mcrt/artistools}{\textsc{artistools}}\footnote{\href{https://github.com/artis-mcrt/artistools/}{https://github.com/artis-mcrt/artistools/}}
\citep{artistools}
were used for data processing and plotting.

\section*{Data Availability}

The data underlying this article will be shared on reasonable request to the corresponding author.



\bibliographystyle{mnras}
\bibliography{astrofritz} 




\appendix

\section{Deviations from reflection symmetry}
\label{sec:deviations_from_symm}

\begin{figure*}
\centering

\includegraphics[width=0.3\textwidth]{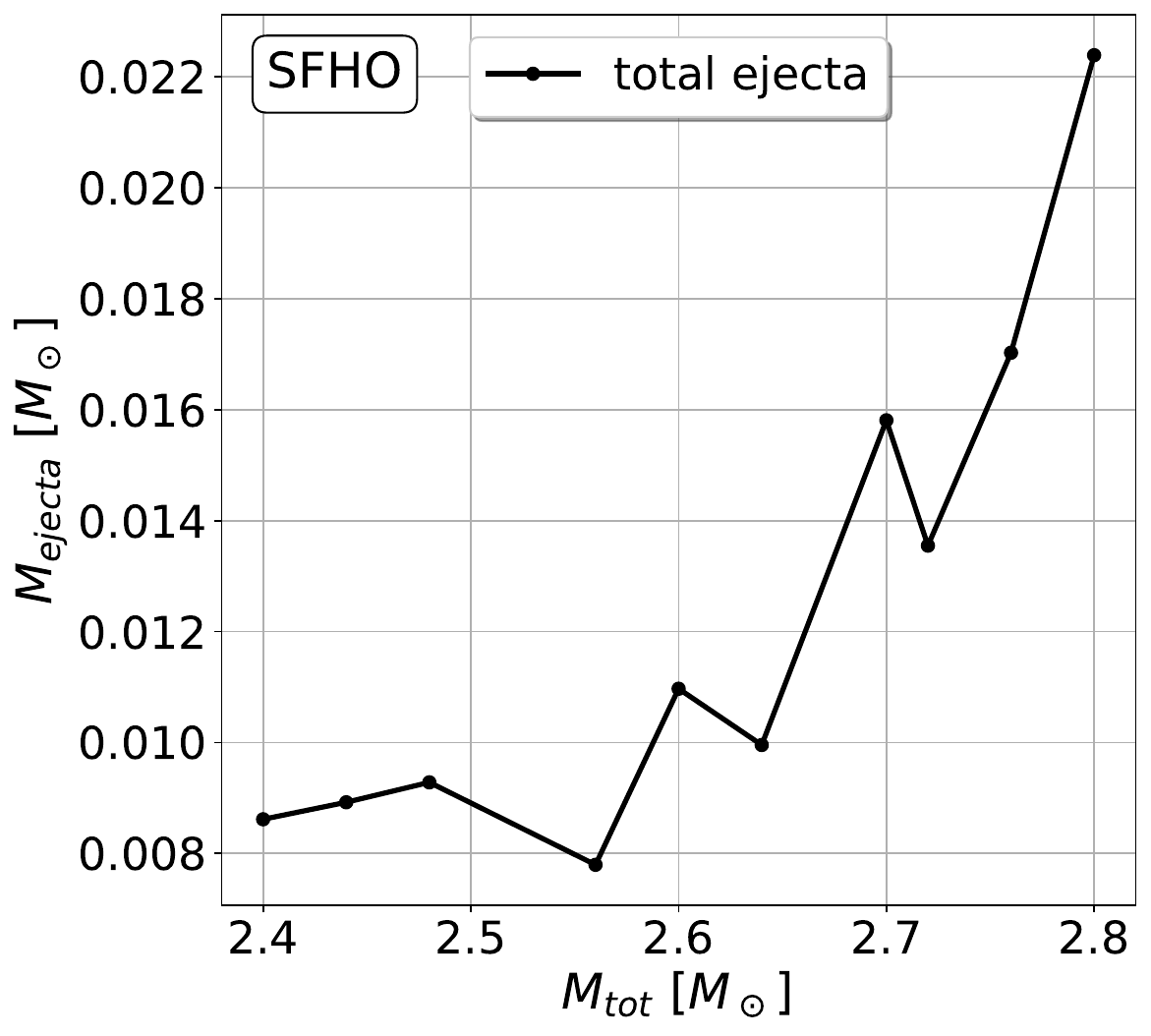}
\includegraphics[width=0.3\textwidth]{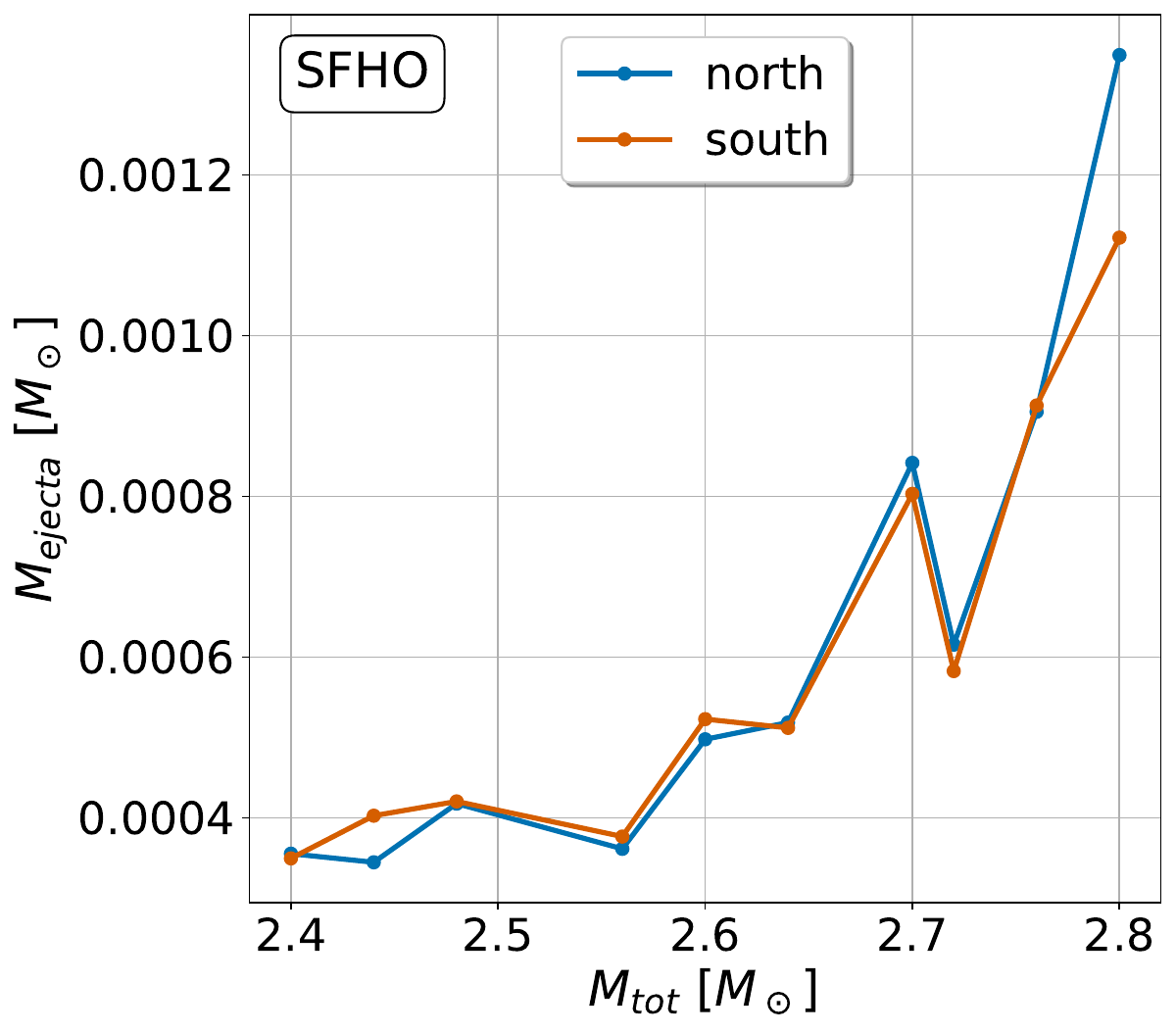}
\includegraphics[width=0.3\textwidth]{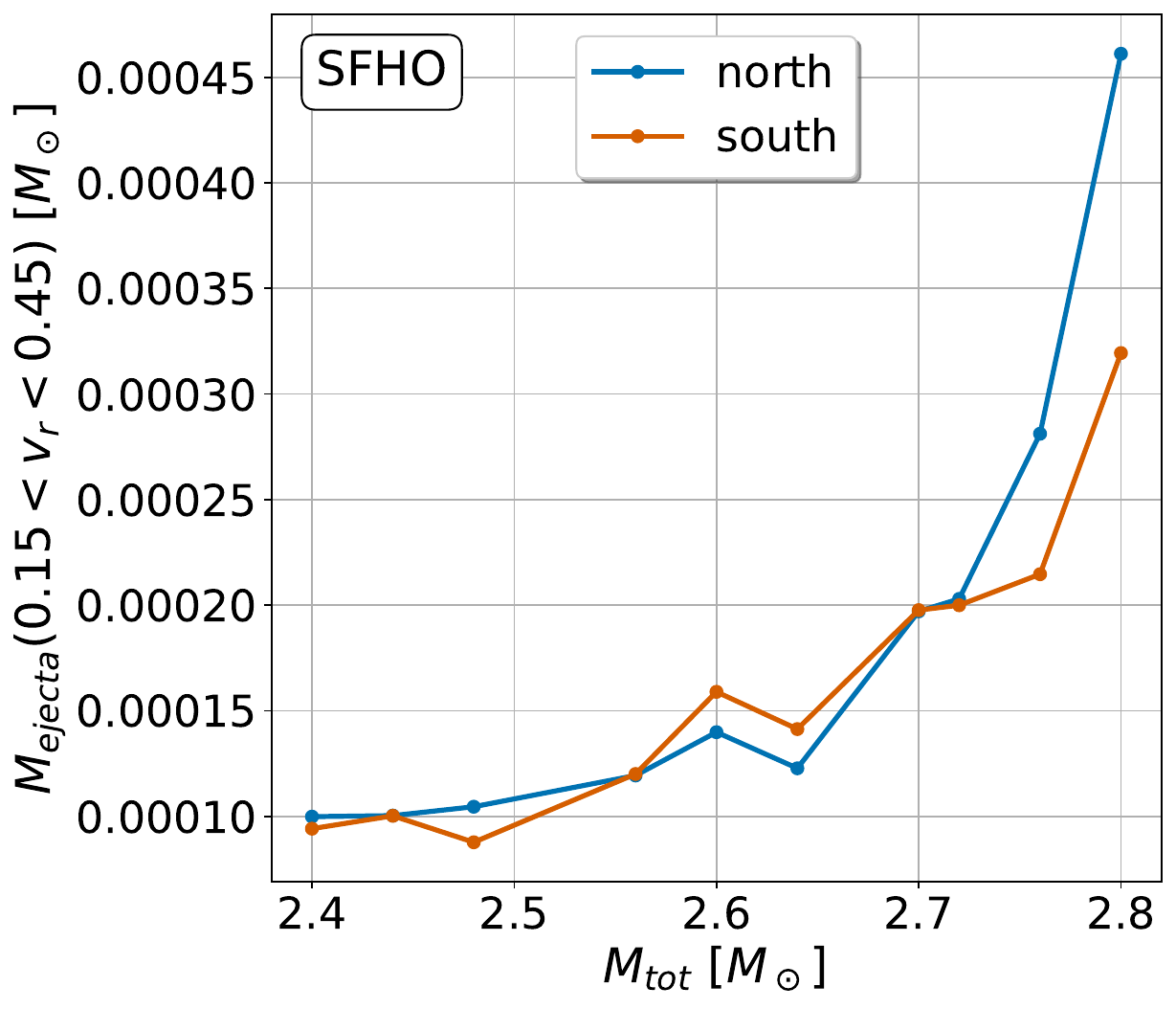}

\includegraphics[width=0.3\textwidth]{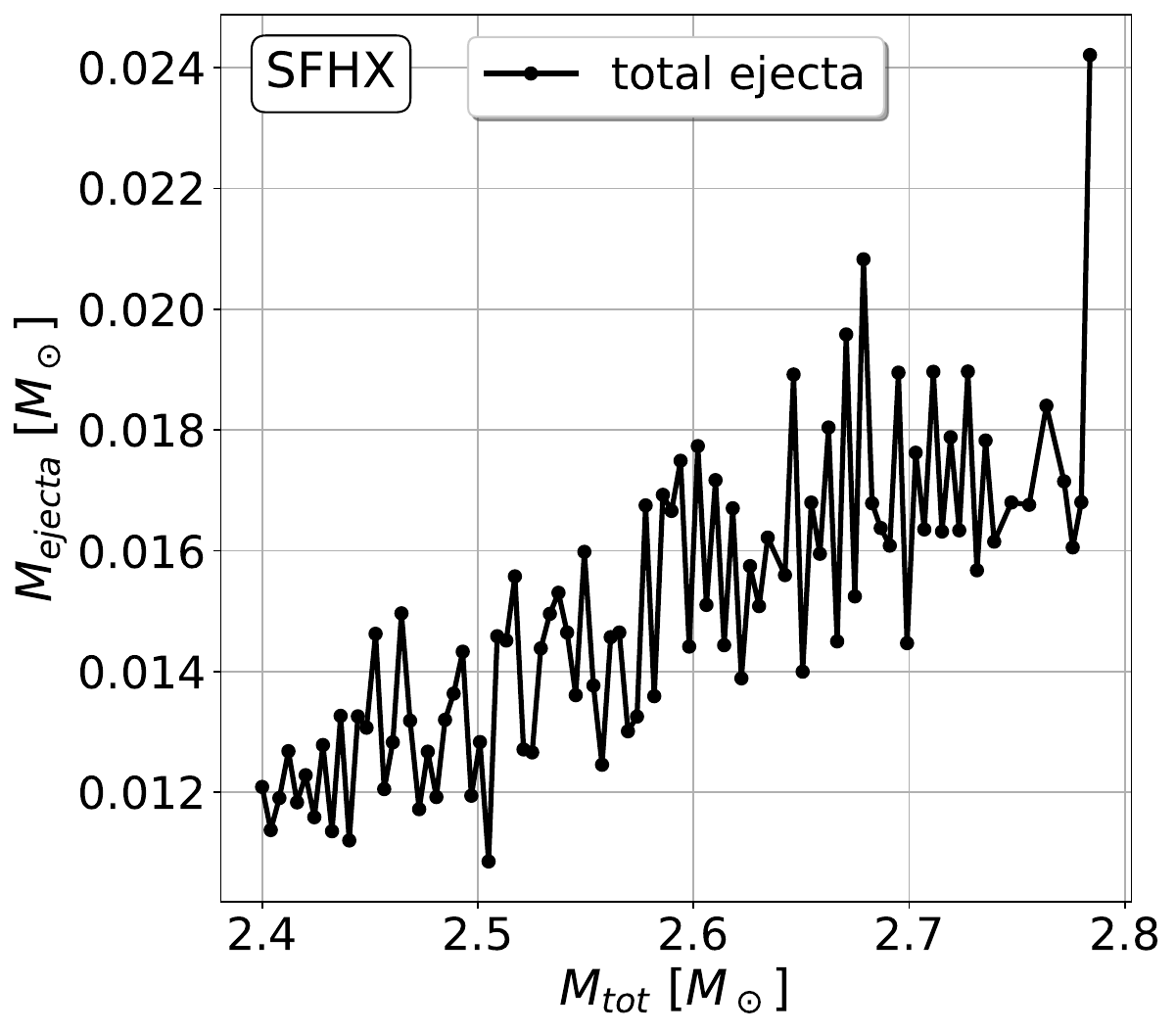}
\includegraphics[width=0.3\textwidth]{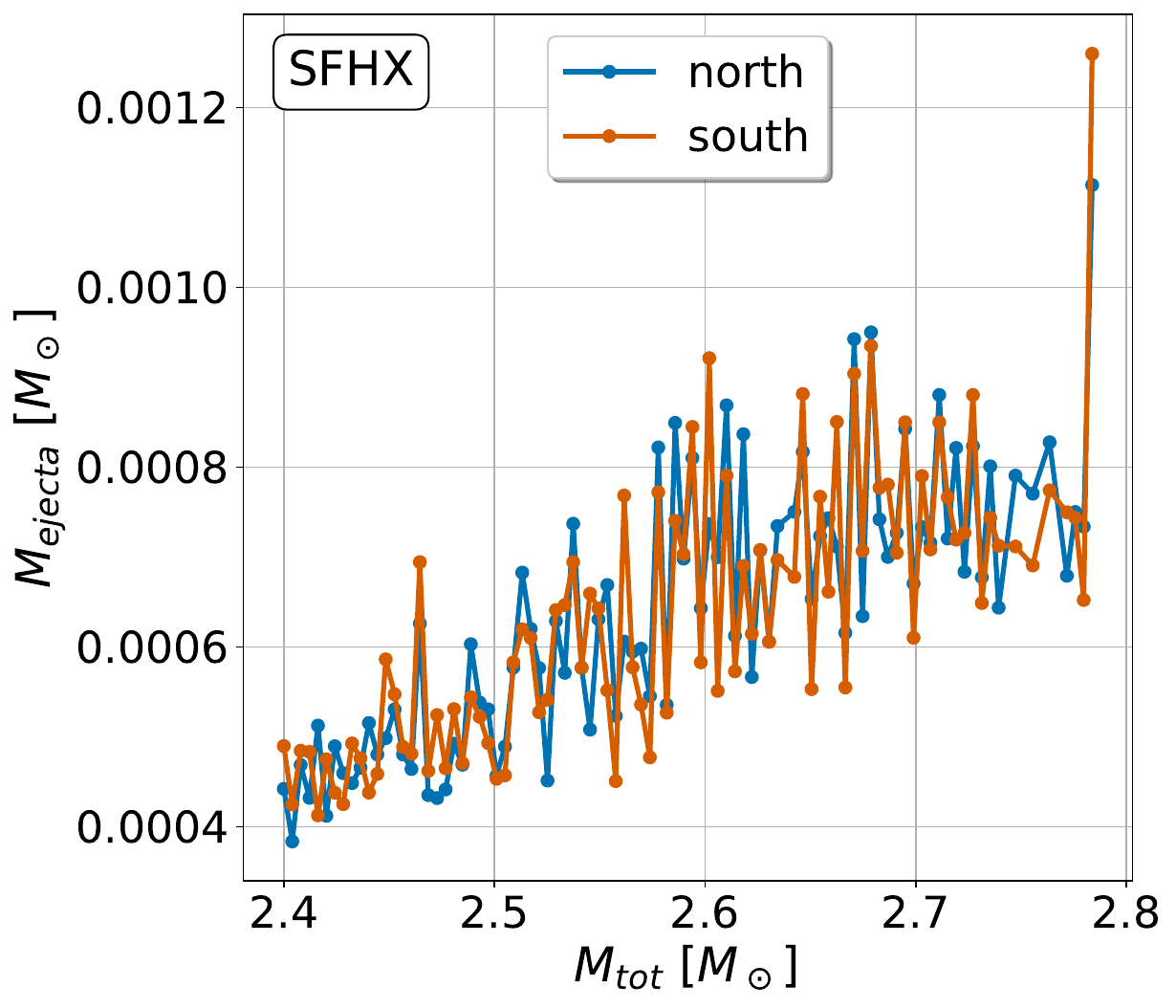}
\includegraphics[width=0.3\textwidth]{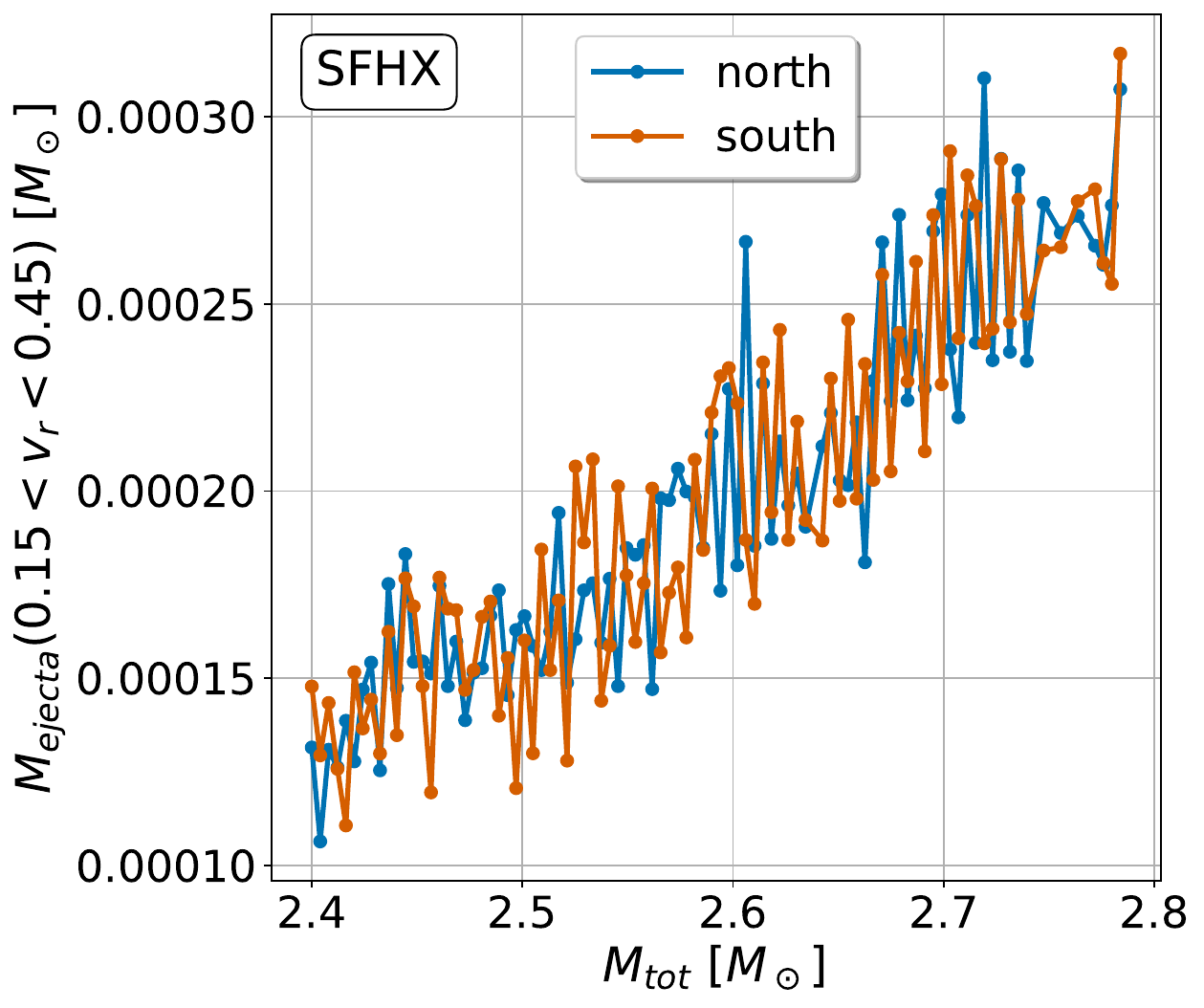}

\caption{Mass of ejecta from neutron star merger simulations with varying total binary mass assuming the SFHo or SFHx equation of state.
The left panels show the total mass ejected by each simulation. The middle panels show the mass ejected into solid-angles at the Northern or Southern pole (with a solid-angle width of cos$(\theta) = 0.2$).
The right panels show the mass within the Northern and Southern solid-angle bins ejected with radial velocities ($v_\mathrm{r}$) in the range \mbox{0.15c < $v_\mathrm{r}$ < 0.45c}.}

\label{fig:models_asymm}
\end{figure*}

The merger simulation used in our \textsc{artis} calculation features a sizeable asymmetry with respect to the orbital plane. For instance, the ejecta masses in the Northern and Southern polar bins differ by a few 10\% (see Figure~\ref{fig:models_asymm}). 
It is not straightforward to assess the extent to which this deviation from reflection symmetry is purely numerical or could be expected to occur in nature.

To better understand the asymmetries in our hydrodynamics tool, we consider a larger set of simulations. We analyse simulations where we vary the total binary mass in small steps for two different equations of state (SFHo and SFHx). The SFHX configurations cover a relatively large total binary mass range with a fine sampling rate. 
In all these simulations neutrinos are not included {and we employ the Wendland SPH kernel \citep{schaback2006a, rosswog2015a}. 
The ejecta masses for a
given model differ from the simulations used in the main text,
which include neutrinos and use the cubic spline kernel \citep{monaghan1985a}}. In Figure~\ref{fig:models_asymm} we show the total ejecta mass and the ejecta mass in the polar bins (as defined in the main text). Additionally, we consider only those fluid elements in the polar directions with a radial velocity in the range $0.15c\leq v \leq 0.45c$. All simulations are analysed 12~ms after merging. Considering that we vary the total binary mass only within a relatively small range, we can expect that bulk properties of the merger would only change slightly. In Figure~\ref{fig:models_asymm} we find that the different near-by models lead to stochastic increase/decrease of the ejecta in the Northern and Southern polar bin. The difference of the mass of polar ejecta are typically about $\sim$10\% in these models implying that the asymmetry of our model used in the main paper may be somewhat on the extreme side of what is a typical symmetry breaking of the simulation code. It is reassuring that there is apparently no systematic trend favouring the ejection in a certain direction.

The magnitude of symmetry breaking visible in Figure~\ref{fig:models_asymm} may be a result of the specific numerical scheme and is likely not informative about the amount of asymmetry in a real merger. Other numerical codes may lead to a different strength of symmetry breaking. Most grid-based merger simulations (presumably) impose reflection symmetry and by surveying the literature we could not easily compare our data to other calculations. The histograms of the ejecta mass at different polar angles in \citet{palenzuela2015a} and \citet{lehner2016a} also show a mild asymmetry of the mass ejection with respect to the orbital plane. It is not obvious to determine if a numerical scheme would overestimate asymmetries or possibly even damp the emergence of deviations from reflection symmetry. At any rate our simulations show that the exact ejecta distribution is apparently very sensitive to small perturbations.

{In addition, we examine the strength of symmetry breaking in simulations with the moving-mesh code \textsc{arepo} \citep{springel2010a}, which was recently extended to describe general relativistic systems \citep{lioutas2024a}. We consider equal-mass mergers for MPA1 \citep{muether1987a, read2009a} ($M_\mathrm{tot}=2.5, 2.7, 2.9~$M$_\odot$), DD2 \citep{hempel2010a, typel2010a} ($M_\mathrm{tot} = 2.7~$M$_\odot$), SFHO \citep{steiner2013a} ($M_\mathrm{tot} = 2.7~$M$_\odot$) and FSU2H \citep{kochankovski2022a} ($M_\mathrm{tot} = 2.8~$M$_\odot$). In all cases, the microphysical EOS is modeled as a zero-temperature barotrope supplemented by a thermal ideal-gas component with $\Gamma_\mathrm{th}=1.75$. We employ a resolution of $m_\mathrm{cell,0}=1.5 \times 10^{-6}~$M$_\odot$ \citep[see section~5.1 in][for more details]{lioutas2024a}. For the system described by MPA1 with $M_\mathrm{tot} = 2.7~$M$_\odot$, we perform additional simulations with lower ($m_\mathrm{cell,0}=3.75 \times 10^{-6}~$M$_\odot$) and higher ($m_\mathrm{cell,0}=6 \times 10^{-7}~$M$_\odot$) resolution as well. We find that these simulations also feature asymmetries of up to some $10\%$ in the ejection along the polar direction. For the setups considered, we do not identify a favoured direction for the polar ejection nor any systematic trend with respect to resolution. Overall, this provides additional support by an independent code that there may be asymmetries with respect to the poles and that they might be of the order $10\%$ in mass.}

It is possible that in nature, as in our simulations, relatively small seed perturbations could be enhanced by the stochasticity of the violent and dynamical merger process and could lead to a considerable asymmetry of the ejecta, implying that a kilonova would appear differently from opposite poles.
Initial perturbations could for instance stem from inhomogenities of the pre-merger magnetic field or from the intrinsic rotation of the neutron stars with one or both spin axes being misaligned with the orbital axis. The latter may introduce a significant mechanism for symmetry breaking even for relatively long rotation periods. If the asymmetries found in our models were representative of actual mergers, the kilonova calculations presented here provide an estimate of the variations that could be expected in nature. Generally, it may not seem unreasonable to expect at least some degree of asymmetry from kilonovae in nature.

\section{Supplemental figures}
\label{sec:supplementalfigs}

\begin{figure*}

\includegraphics[width=0.24\textwidth]{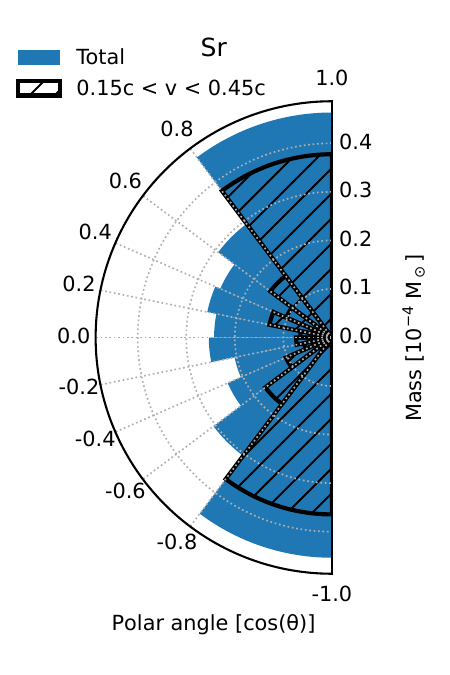}
\includegraphics[width=0.24\textwidth]{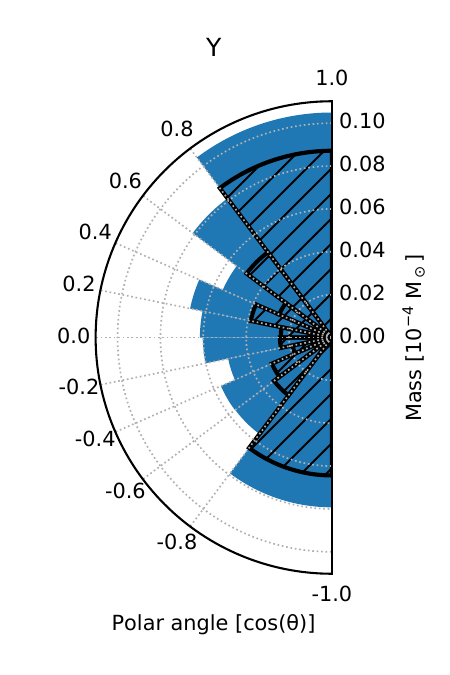}
\includegraphics[width=0.24\textwidth]{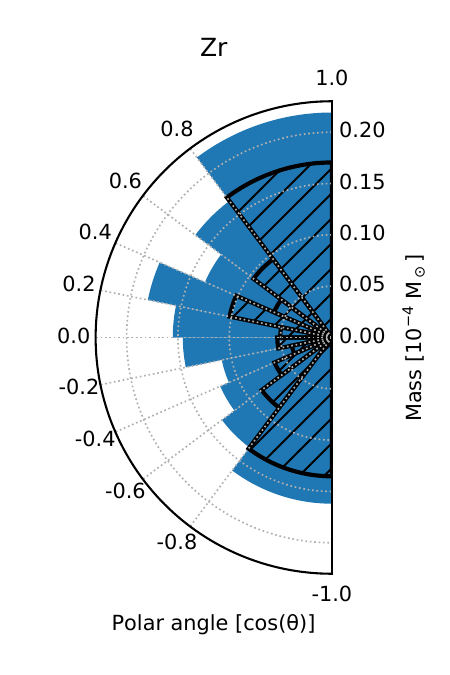}
\includegraphics[width=0.24\textwidth]{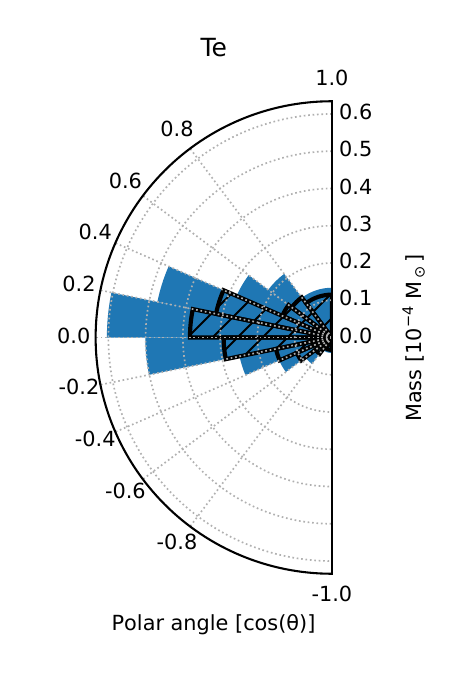}

\includegraphics[width=0.24\textwidth]{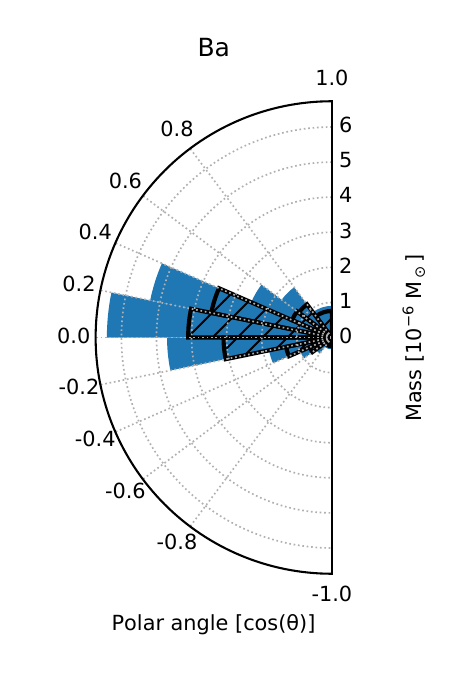}
\includegraphics[width=0.24\textwidth]{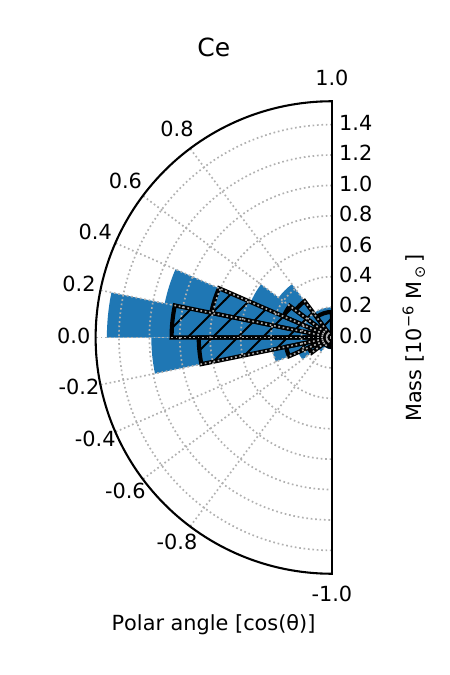}
\includegraphics[width=0.24\textwidth]{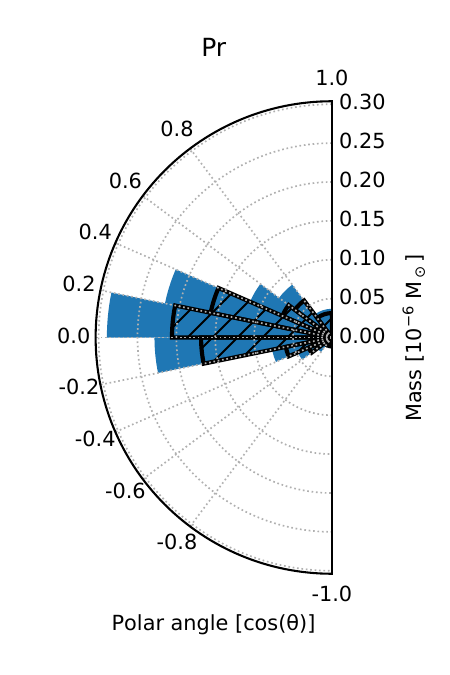}
\includegraphics[width=0.24\textwidth]{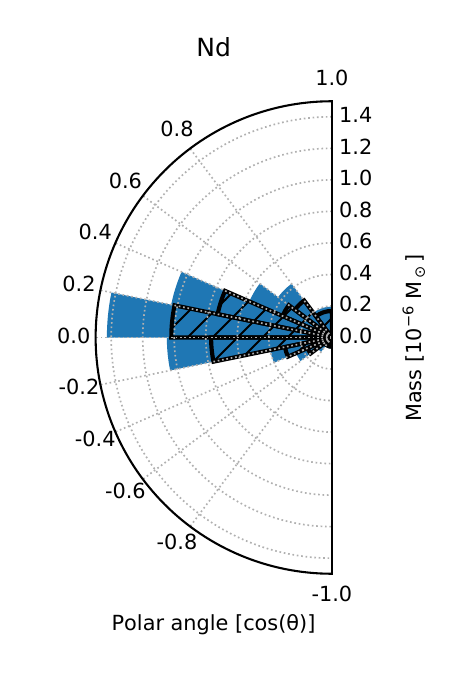}

\caption{
Mass ejected per polar angle of select elements synthesised in the model.
Note the mass scale for the lower plots is 10$^2$ times lower than for the upper plots.
The ejecta mass in the velocity range $0.15c < v < 0.45c$ is shown by the black lines, indicating the ejecta mass which is approximately within the line forming region in the first day after the merger.
}

\label{fig:mass-histograms-elements}
\end{figure*}

\begin{figure*}
\centering

\includegraphics[width=0.8\textwidth]{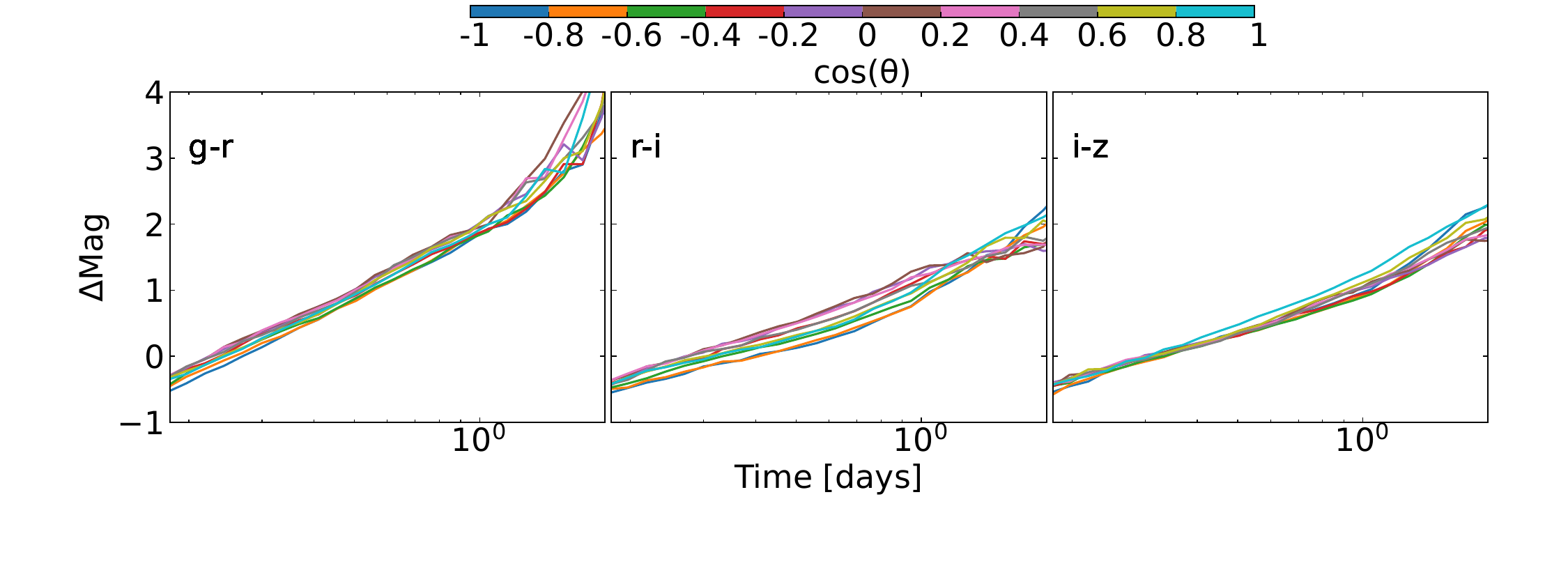}

\caption{Direction dependent colour evolution, showing that in all lines of sight the light curves initially show blue colours and evolve to redder colours with time. The higher abundance of lanthanide rich material near the equator does not lead to significantly redder colours in the light curves.}

\label{fig:colourevolution}
\end{figure*}

The mass distribution of select elements is shown in Figure~\ref{fig:mass-histograms-elements}, demonstrating that heavier,
higher opacity elements are predominantly produced near the equator.
This higher abundance of heavy elements does not lead to a significantly redder colour in the light curves at the equator, as demonstrated by Figure~\ref{fig:colourevolution}.
In all directions, the light curves initially exhibit blue colours and evolve to redder colours over time.

\bsp	
\label{lastpage}
\end{document}